\documentclass[useAMS,usenatbib]{mn2e}
\usepackage{graphicx} 
\usepackage{times}
\usepackage{epsfig}
\usepackage{rotating}
\usepackage{sidecap}
\usepackage{longtable}
\usepackage{lscape}
\usepackage{rotating}
\usepackage{amssymb,xcolor}

\title[Geometrical constraints on black hole timing properties]{Geometrical constraints on the origin of timing signals from black holes}

\author[S. Motta et al.]{S.E. Motta$^{1}$, P. Casella$^2$, M. Henze$^{1}$, T. Mu\~noz-Darias$^3$, A. Sanna$^4$, R. Fender$^3$, T. Belloni$^5$ \\
$^{1}$ESAC, European Space Astronomy Centre, Villanueva de la Ca\~nada, E-28692 Madrid, Spain\\
$^{2}$INAF-Osservatorio Astronomico di Roma, Via Frascati 33, I-00040, Monteporzio Catone, Italy\\
$^{3}$University of Oxford, Department of Physics, Astrophysics, Denys Wilkinson Building, Keble Road, Oxford OX1 3RH, UK\\
$^{4}$Dipartimento di Fisica, Universit\'a degli Studi di Cagliari, SP Monserrato-Sestu km 0.7, 09042 Monserrato, Italy \\
$^{5}$INAF-Osservatorio Astronomico di Brera, Via E. Bianchi 46, I-23807 Merate, Italy\\
}
\begin{document}
\maketitle
\begin{abstract}

We present a systematic study of the orbital inclination effects on black-hole transients fast time-variability properties. We have considered all the black-hole binaries that have been densely monitored  by the Rossi XTE satellite. We find that the amplitude of low-frequency quasi periodic oscillations (QPOs) depends on the orbital inclination. Type-C QPOs are stronger for nearly edge-on systems (high inclination), while type-B QPOs are stronger when the accretion disk is closer to face-on (low inclination). Our results also suggest that the noise associated with type-C QPOs is consistent with being stronger for low-inclination sources, while the noise associated to type-B QPOs seems inclination independent. 
These results are consistent with a geometric origin of the type-C QPOs - for instance arising from relativistic precession of the inner flow within a truncated disk -  while the noise would correspond to intrinsic brightness variability from mass accretion rate fluctuations in the accretion flow.
The opposite behavior of type-B QPOs - stronger in low inclinations sources -  supports the hypothesis that type-B QPOs are related to the jet, the power of which is the most obvious measurable parameter expected to be stronger in nearly face-on sources. 
\end{abstract}



\section{Introduction}

Quasi-periodic oscillations (QPOs) were discovered several decades ago in the  X-ray flux emitted from accreting black hole (BH) binaries and have been observed in many systems. 
In a Fourier power density spectrum (PDS) they take the form of relatively narrow peaks and appear together with different kinds of broad-band noise (e.g. \citealt{Takizawa1997}, \citealt{Casella2005}, \citealt{Motta2012}). It is now clear that QPOs are a common characteristic of accreting BHs and they have been observed also in neutron stars (NS) binaries (e.g. \citealt{VdK1989}, \citealt{Homan2002a}, \citealt{Belloni2007}), in cataclysmic variable (see e.g. \citealt{Patterson1977}) in the so-called \textit{ultra luminous X-ray sources} (possibly hosting intermediate-mass BHs, e.g.  \citealt{Strohmayer2003a}, \citealt{Strohmayer2003a}) and even in Active Galactic Nuclei (AGNs, e.g. \citealt{Gierlinski2008}, \citealt{Middleton2010}). 

Low-frequency QPOs (LFQPOs), with frequencies ranging from a few mHz to $\sim$20 Hz were first observed with \emph{Ariel 6} in the BH binary GX 339-4 (\citealt{Motch1983}) and observations with \emph{Ginga} provided the first indications for the existence of multiple types of LFQPOs (see e.g. \citealt{Miyamoto1991} and \citealt{Takizawa1997}). Three main types of LFQPOs, type-A, type-B, and type-C\footnote{Type-C QPOs are by far the most common type of QPOs observed in BH binaries. Their amplitude is usually large and they are characterized by a variable frequency ranging the 0.1-30 Hz interval. Type-B QPOs are less common than the type-C QPOs, they are usually observed along the transition form hard to soft in transient BH binaries with frequencies around $\sim$6Hz. Among LFQPOs, type-A QPOs are the least common of all. They usually appear in the soft states of transient BH binaries as broad and weak peaks.}, originally identified in the PDS of XTE J1550-564 (see e.g. \citealt{Wijnands1999}; \citealt{Homan2001}; \citealt{Remillard2002}, \citealt{Casella2005}, \citealt{Motta2011a}), have been seen in several sources. These are distinct from the high-frequency QPOs, found at frequency up to $\sim$450Hz, which we do not consider in this work (but see \citealt{Belloni2012} for a review).

The origin of LFQPOs is still unclear and there is no consensus about their physical nature, although their study provides a valuable way to explore accretion around accreting compact objects.
Several models have been proposed to explain the origin and the evolution of LFQPO in X-ray binaries. Some of them invoke the effects of General Relativity (GR) (e.g. \citealt{Stella1999a}, \citealt{Ingram2009}, \citealt{Motta2014}), while others are based on different kinds of instabilities (e.g. \citealt{Titarchuk1999}, \citealt{Tagger1999}, \citealt{Lamb2001}). On the other hand, several doubts still surround the origin of type-B QPOs, for which no comprehensive model has been proposed. However, it has been speculated that type-B QPOs are associated with the strong relativistic jets that occurs in BH binaries (BHB) during specific state transitions (\citealt{Fender2009}, \citealt{Miller-Jones2012}). 

It has been known for a long time that inclination strongly affects the observed properties of AGNs (see e.g.  \citealt{Antonucci1993}, \citealt{Urry1995}, \citealt{Risaliti2002}, \citealt{Bianchi2012}). Over the last years, it has become increasingly clear that the same is true for Galactic accreting BHBs.
\cite{Ponti2012} found strong evidences that the accretion disk winds observed in the radio quiet soft states of BHBs have an equatorial geometry with opening angles of a few degrees and therefore can only be observed in sources where the disc is inclined at a large angle \textit{i} to the line of sight (\textit{high-inclination sources}, as opposed to \textit{low-inclination sources}, where the orbital plane is closer to perpendicular to the line of sight). 
More recently, the results by \cite{Munoz-Darias2013} supported the hypothesis that the inclination modifies the q-shaped tracks that BHB in outburst display in a hardness-intensity diagram (HID, \citealt{Homan2001}), which can be at least partially explained by considering inclination-dependent relativistic effects on the accretion disc. Finally, \cite{Corral-Santana2013} have found that obscuration effects similar to those observed in AGN, can be relevant in very high inclination BHBs. However, \cite{Gallo2003} and \cite{Soleri2011} noted that there is no evidence that the hard state radio luminosity is a function of inclination.  

In this work, we use data collected by the Rossi X-ray Timing Explorer (RXTE)/Proportional Counter Array (PCA) satellite to analyse the effects of inclination on the fast time-variability properties of BHBs. 
	\begin{table*}	
	\renewcommand{\tabcolsep}{0.18cm}															
	\centering																	
	\caption{List of black hole transients and outbursts included in this work. In the column \textit{comments} we report some informations about the behaviour of the sources relevant to distinguish between high and low inclination systems. With \textit{high inclination evolution (High-IE)}, \textit{intermediate inclination evolution (Int.-IE)} and \textit{low-inclination evolution (Low-IE)} we refer to high, intermediate and low disk temperatures, respectively, as reported in \citealt{Munoz-Darias2013}. As discussed by these authors, the differences in the disk temperatures  can be largely ascribed to the inclination of the disk to the line of sight.With the term \textit{spikes} we refer to flux spikes visible in both the lightcurve and HIDs of most high-inclination systems (see \citealt{Munoz-Darias2013}). The term \textit{winds} or \textit{dipping} in the \textit{comments} column indicate that equatorial winds or absorption dips, respectively, have been reported for that source. }\label{tab:sources}	
	\begin{tabular}{c c c c c c c c}		
		\hline																	
	System	&	Outburst	&	Inclination$^e$ ($^{\circ}$)	&		Comments	&	Ref.		&	Type-A	&	Type-B	&	Type-C	\\
		\hline																	
	Swift J1753.5-01	&	2005-2010	&	$\sim$ 40--55 	&		Failed outburst $^{(d)}$	&	0		&		&		&	32	\\
	4U 1543-47	&	2002	&	$20.7 \pm 1.5$	&		Low-IE;	&	1		&	2	&	3	&	11	\\
	XTE J1650-500	&	2001	&	$>$ 47	&		Low-IE;	&	2		&		&	1	&	25	\\
	GX 339-4	&	2002, 2004, 2007, 2010	&	$\geq$ 40$^{(c)}$	&		Low-IE; 	&	3		&	4	&	23	&	54	\\
	XTE J1752-223	&	2009	&	$\leq$ 49$^{(a)}$	&		Low-IE;	&	4		&		&	2	&	4	\\
	XTE J1817-330	&	2006	&		&		Low-IE;	&			&	2	&	9	&	2	\\
	\hline 																	
																		
	XTE J1859+226	&	1999	&	$\geq$ 60$^{(b)}$	&		Int.-IE (spikes)	&	5		&	5	&	19	&	24	\\
	MAXI J1543-564	&	2011	&		&		Spikes;	&			&		&		&	5	\\
																		
	\hline																	
	XTE J1550-564	&	1998, 2000	&	$74.7 \pm 3.8$	&		high-IE (spikes); dipping;	&	6,7		&	1	&	18	&	48	\\
	4U1630-47	&	2002, 2003, 2004, 2005	&		&		high-IE (spikes); dipping; winds;	&	8, 9		&		&	6	&	19	\\
	GRO J1655-40	&	1996, 2005	&	$70.2\pm 1$	&		high-IE (spikes); dipping; winds;	&	8, 10		&		&	4	&	50	\\
	H1743-322	&	2003, 2004, 2008 (Jan. and Oct.), 	&	$75\pm3$ $^{(a)}$	 &		high-IE (spikes); dipping; winds;	&	11, 12, 13		&		&	42	&	108	\\
		&	2009, 2010 (Jan. and Aug.)  	&		&			&			&		&		&		\\
	MAXI J1659-152	&	2010	&		&		Spikes; dipping;	&	14		&		&	8	&	40	\\
	XTE J1748-288	&	1998	&		&		Spikes; dipping;	&			&		&		&	7	\\
\hline	
	\end{tabular}
	\begin{flushleft}																	
	(a) Assuming that the radio jet is perpendicular to the accretion disk.\\	
	(b) Inclination $\sim$ 60 degrees if accretion disc does not contribute to the optical luminosity in quiescence.\\																	
	(c) The constrain is placed assuming that the mass of the BH should not be larger than 20 M$_{\odot}$.\\			(d) Failed outburst (see \citealt{Soleri2013}).\\
	(e) Inclination measurements come either form the X-rays or from multi-wavelength observations (moslty optical and radio).\\
																		
	\textsc{REFERENCES:} 																	
	(0) \cite{Neustroev2014} ;~(1) \citet{Orosz2002};~(2) \citet{Orosz2004};~(3) \citet{Munoz-Darias2008};~(4) \cite{Miller-Jones2011};		~(5) \citet{Corral-Santana2011};~(6) \citet{Orosz2011}; ~(7) \citet{Homan2001}; ~(8) \citet{Kuulkers1998}; ~ (9) \citet{Tomsick1998};~(10) \citet{Greene2001};	~(11) \citet{Corbel2005};	~(12) \citet{Steiner2012}; ~(13) \citet{Homan2005b}; ~(14) \citet{Kuulkers2013};															\end{flushleft}	
	\end{table*}	
%
%
%
\section{Observations and data analysis}\label{sec:observations} 

The main goal of this work is to perform a systematic study on the effect of the orbital inclination on the most common QPOs observed in BHs (i.e. type-C and type-B QPOs) and their associated noise. 

To do so, we assume that there is no intrinsic difference between low inclination and high inclination systems. Therefore, any systematic difference observed  between the QPO properties of systems viewed at different inclinations has to be mainly due to this parameter.  We base our assumption on the fact that the outburst evolution and the macroscopic properties (e.g. orbital parameters, see \citealt{Munoz-Darias2013} where a similar source selection is performed) of these systems do not show any obvious difference. 

\subsection{The Sample}\label{sec:sample}

For our analysis we considered only the sources that have shown LFQPOs. We only investigate those that have been densely monitored by RXTE, in order to maximize the chances of observing high-inclination features (i.e. X-ray absorption dips or eclipses), if present. This is because accurate inclination measurements are not generally available for most of the sources (see \citealt{Casares2014})\footnote{Note, however, that the accurate inclination measurements available for a few sources (see Tab. \ref{tab:sources}) mostly come from optical spectro-phofometric observations (see Appendix \ref{App:AppendixB}).}. We therefore used the presence/absence of  absorption dips as the main discriminant to separate high and low inclination sources (see also Appendix \ref{App:AppendixA} and \ref{App:AppendixB}). Note that transient BHBs usually have relatively short orbital periods (from a few days down to a few hours, see e.g. \citealt{Casares2014}), therefore absorption dips can appear quite often and it is in principle reasonably easy to detect them with a good observation coverage. This simplistic approach was applied in two previous works where it has been shown that the sources of both groups present, depending on the orbital inclination, different behaviours when looking at the spectral properties (i.e. colours and disk temperatures, \citealt{Munoz-Darias2013}) and at the presence/absence of winds (\citealt{Ponti2012}). 

Our sample includes the following 14 sources: Swift J1753.5-01, 4U 1543--47, XTE J1650-500, GX 339-4, XTE J1752-223, XTE J1817-330, 4U 1630-47, GRO J1655-40, H1743-322, MAXI J1659-15, XTE J1748-288, XTE J1550-564, XTE J1859+226, MAXI J1543-564  (see Tab. \ref{tab:sources}). XTE J1859+226 and MAXI J1543-564 have been treated separately because they cannot be unambiguously placed in either the high-inclination or the low-inclination group (see Appendix \ref{App:AppendixB}). We refer to them as intermediate-inclination sources.  

We excluded from this work XTE J1118+480, XTE J1652--453 and XTE J1720--318, GS 1354--64\footnote{Additionally, GS 1354--64, has been observed by RXTE only 7 times.} because they did not display any significant QPO during the outbursts covered by RXTE. We also excluded GRS 1915+105 (see e.g. \citealt{Fender2004a}, \citealt{Soleri2008}) and IGR J17091-3624 (e.g. \citealt{Altamirano2011}) because of their unusual outburst evolution, which is different to that of the other sources (i.e. they did not display a clear q-shaped HID). 


The classification above results in a grouping where high and low inclination roughly mean inclination angle larger and smaller than $\sim$ 65-70$^{\circ}$, respectively. This value, however, must be taken with caution as the exact threshold for high-inclination depends on the characteristic of each system, such as mass ratio between black hole and companion star, orbital separation, dimension of the accretion disk and of the bulge that forms were the material from the companion star hits the accretion disk. This topic is addressed in detail  in Appendix \ref{App:AppendixA}.

\subsection{Data Analysis}

We examined all the RXTE archival observations of the sources in our sample.
For each observation we computed power spectra from RXTE/PCA data using custom software under \textsc{IDL}\footnote{GHATS, http://www.brera.inaf.it/utenti/belloni/GHATS\_Package/Home.html} in the energy band 2-26 keV (absolute PCA channel 0 to 62). We used 128s-long intervals and a Nyquist frequency of 1024 Hz. We averaged the PDS and subtracted the contribution due to Poissonian noise (see \citealt{Zhang1995}). The PDS were normalized according to \cite{Leahy1983} and converted to square fractional rms (\citealt{Belloni1990}).  

We selected for our analysis only observations where a somewhat narrow (quality factor\footnote{$Q = \nu_{centroid}/FWHM$,  where $\nu_{centroid}$ is the centroid frequency of the QPO fitted with a Lorentzian and FWHM its full width half maximum)} Q $>$ 2)  low-frequency ($<$ 50 Hz) feature was identifiable on top of flat-top or power-law shaped noise components in the PDS. 
PDS fitting was carried out with the {\sc xspec} package by using a one-to-one energy-frequency conversion and a unit response. Following \cite{Belloni2002}, we fitted the noise components with a number of broad Lorentzian shapes. LFQPOs are well-described by a variable number of narrow Lorentzians depending on the presence of harmonic peaks. When more than one peak was present, we identified the fundamental based on the QPO evolution along the outburst.
Based on the results of the fitting, we excluded from the analysis non-significantly detected features (significance\footnote{The significance of QPOs are given as the ratio of the integral of the power of the Lorentzian used to fit the QPO divided by the negative 1-sigma error on the integral of the power.} $\leq$ 3$\sigma$). 
We measured the rms as the square root of the integral over all the available frequencies of each component in the fit to the PDS. This means that the rms values we reported are measured between 1/128 Hz and 1024 Hz. For each PDS, we measured the fractional rms of each component of the spectrum (i.e. each Lorentzian used to fit the spectrum).  Whenever a QPO is formed by more than one harmonic peak, we measured the rms of the QPO adding in quadrature the rms of  the harmonic peaks.  The rms of the noise comes from all the remaining components. The total rms is measured adding in quadrature the rms of all the components of the PDS. 
We report in Table \ref{tab:parameters} the rms of the QPO, the rms of the noise and the total rms for each observation. 

We classified the LFQPOs following the method outlined by \citealt{Motta2012} (based on \citealt{Casella2005}), dividing them into type-A, type-B and type-C QPOs. We collected a total of 429 type-C QPOs, 135 type-B QPOs and 14 type-A QPOs (two from 4U 1543--47, two from XTE J1550-564, two from XTE J1817-330, four from XTE J1859+226 and four from GX 339-4). Due to the very low number of detections, we decided to exclude type-A QPOs from this work. 
In order to differentiate between the rms of the Type-B/C QPO, the rms of the noise associated with a Type-B/C QPO and the total rms of a PDS where we detected a type-B/C QPO, we will refer to type-B/C QPO rms, type-B/C noise rms and type-B/C total rms, respectively. 
\begin{figure*}
\centering
\hfill
\includegraphics[width=0.45\textwidth]{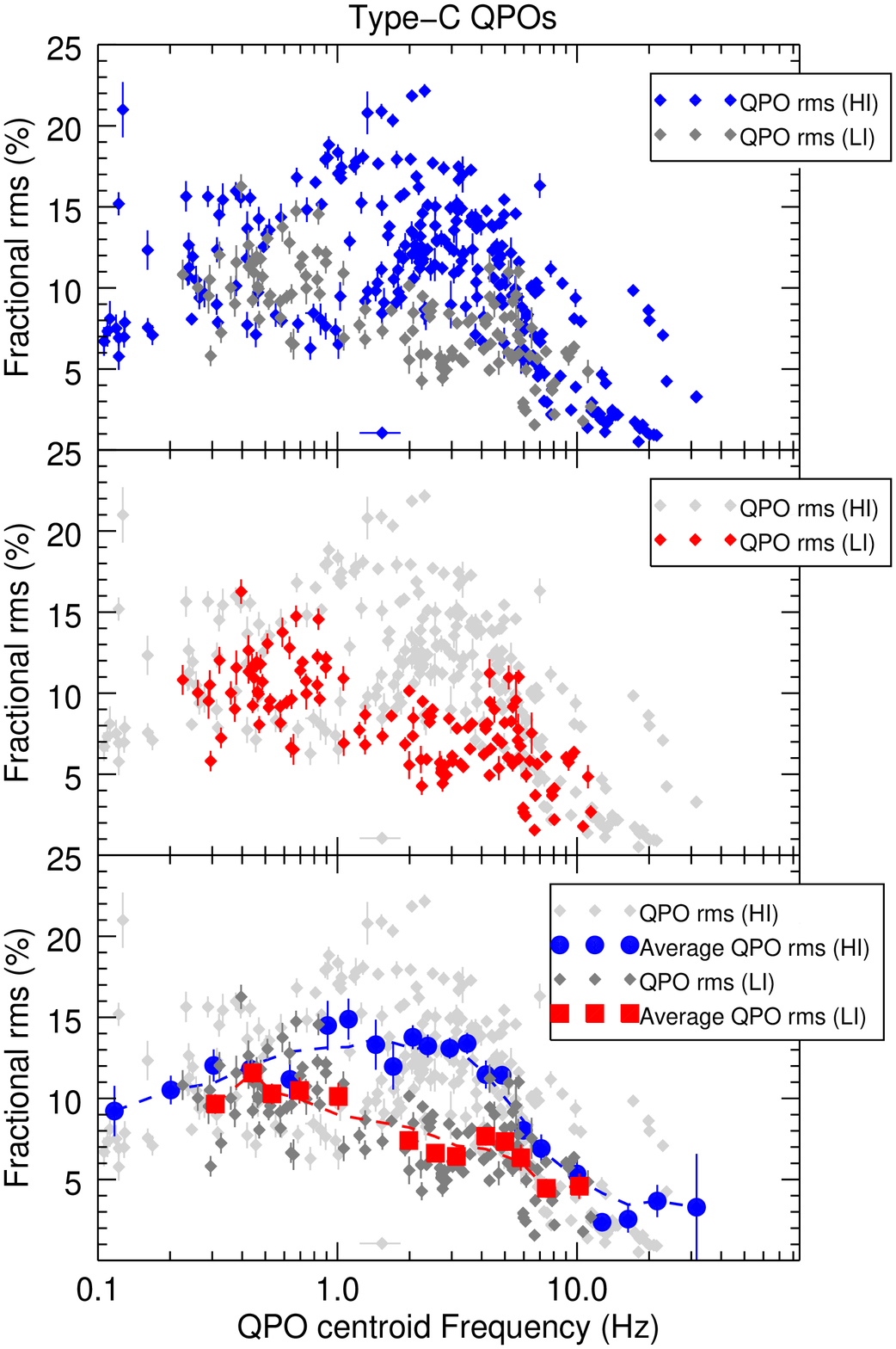}
\hfill
\includegraphics[width=0.45\textwidth]{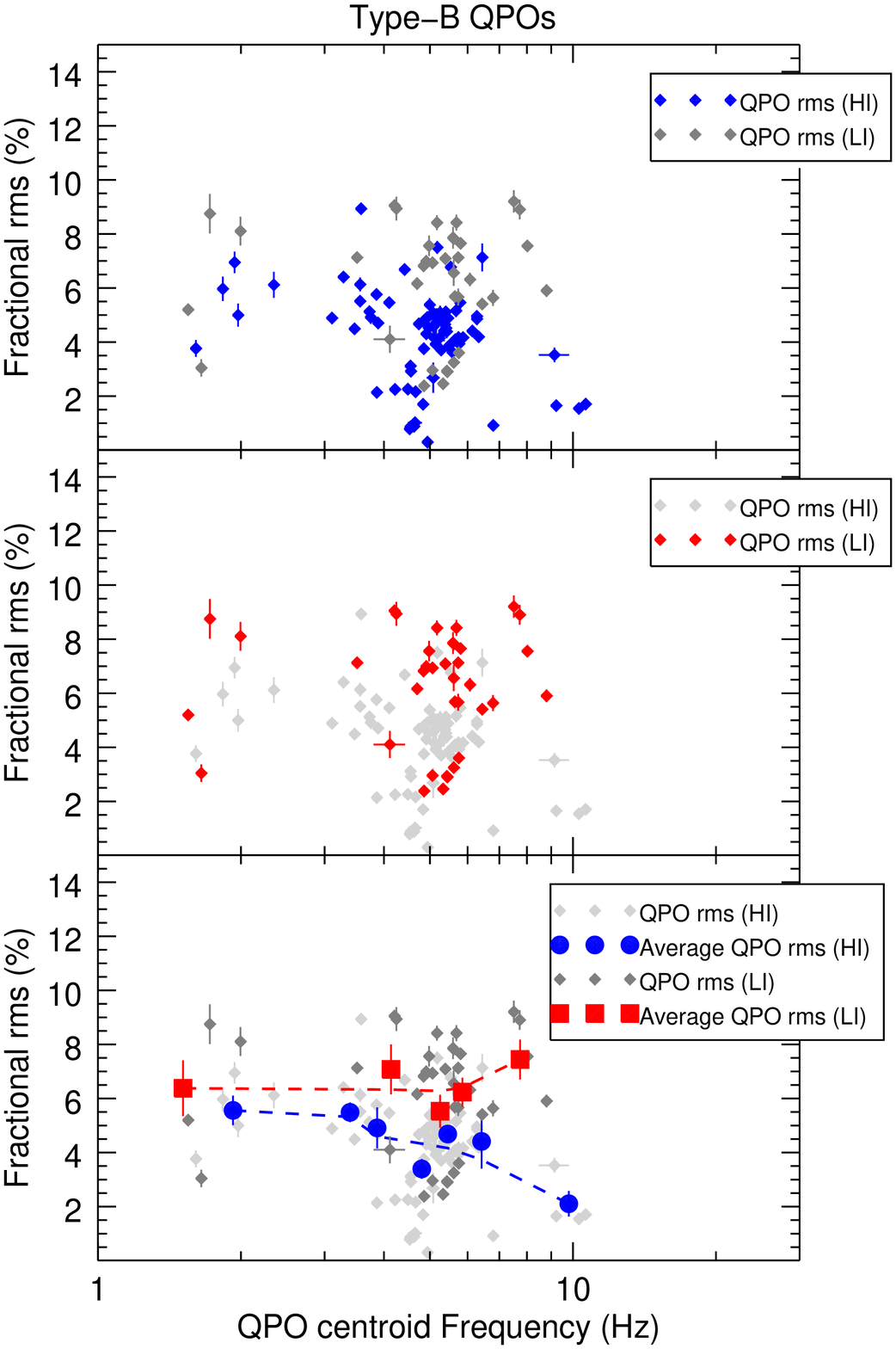}
\hfill
\caption{QPO rms as a function of the QPO centroid frequency for Type-C QPOs (left figures) and Type-B QPOs (right figures). For each figure we plot as follows. In the top panel we show data from high-inclination (HI) sources (blue) and from low-inclination (LI) sources (dark grey). In the middle panel we plot  the data from low-inclination sources (red) and from high inclination sources (light grey). In the bottom panel light grey and dark grey points are the same as in the top and middle panel, the blue circles correspond to QPO detected in high-inclination sources, the red squares correspond to QPOs detected in low-inclination sources. The red squares and blue circles are \textquoteleft average\textquoteright\ points and have been obtained by applying a logarithmic rebin in frequency to the grey points. The dashed blue and red lines are smooth fits to the blue and red points respectively, for visualising. Here we did not separate the sources, but we only distinguish between high- and low-inclination ones. Note that the axes are scaled differently for type-C and -B QPOs.}\label{fig:QPO_vs_freq}
\end{figure*}
\begin{figure*}
\centering 
\hfill
\includegraphics[width=0.45\textwidth]{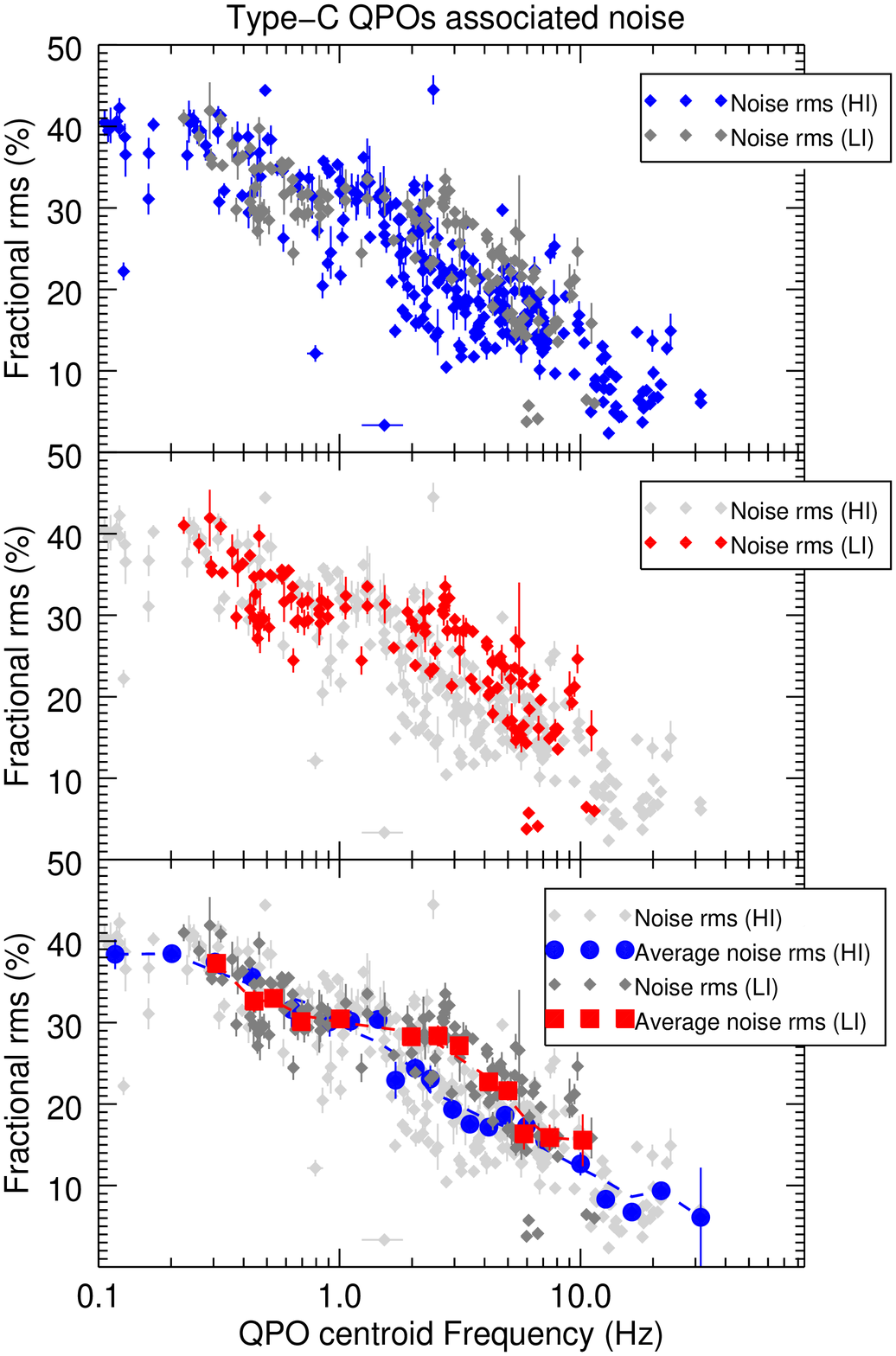}
\hfill
\includegraphics[width=0.45\textwidth]{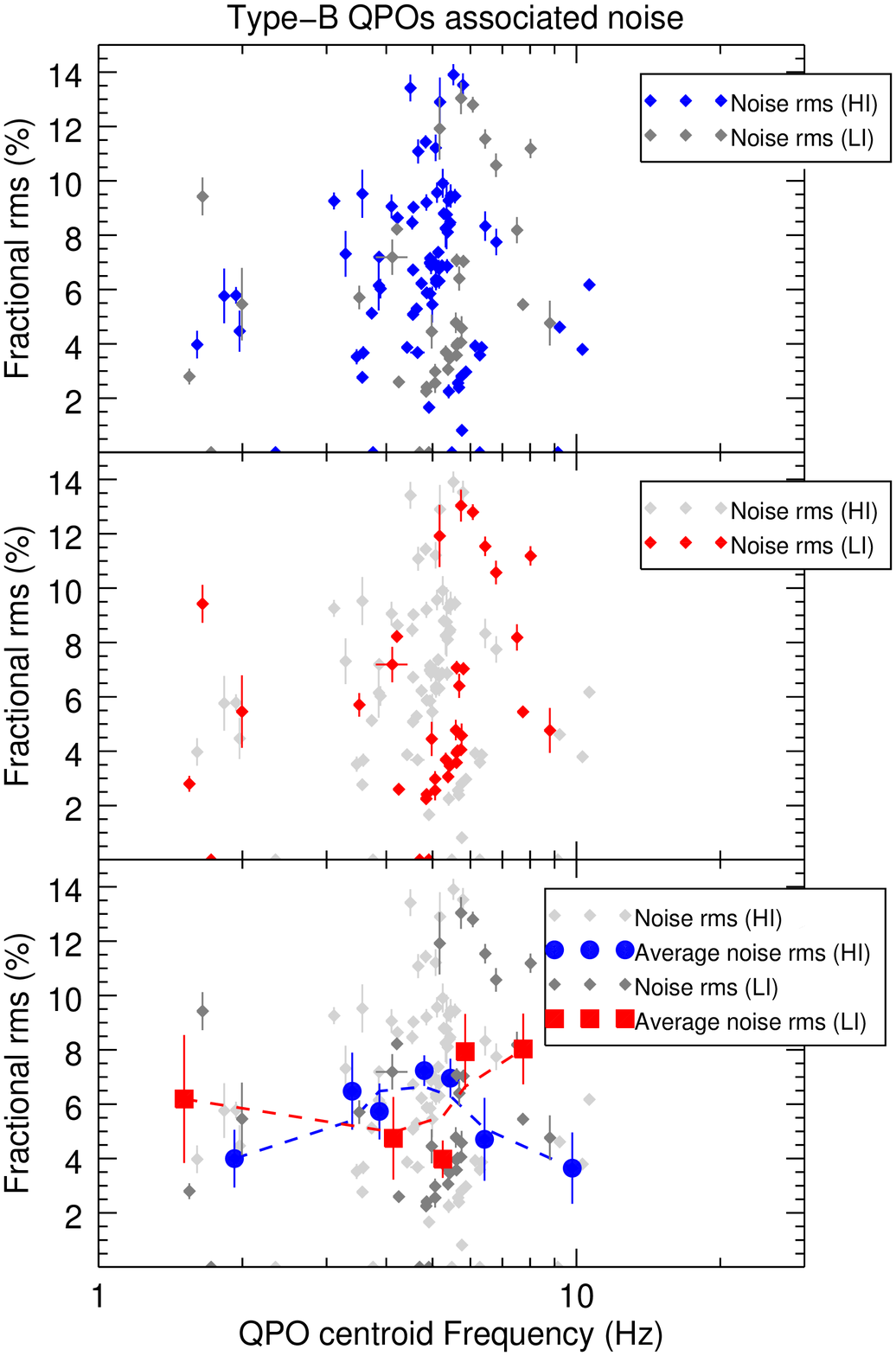}
\hfill
\caption{Rms of the noise associated with the QPOs in Fig. \ref{fig:QPO_vs_freq} as a function of the QPO centroid frequency. Color code is the same for Fig. \ref{fig:QPO_vs_freq}. The average points have been produced as described in Fig. \ref{fig:QPO_vs_freq}.}\label{fig:noise_vs_freq}
\end{figure*}
%
\section{Results}\label{Sec:results}

After classifying QPOs, we plotted the QPO rms and the noise rms as a function of the QPO centroid frequency, dividing the sources in high inclination and low inclination: we do not separate the different sources in order to give a better idea of the data general trend. The results are shown in Fig. \ref{fig:QPO_vs_freq} and \ref{fig:noise_vs_freq}. Blue and red points always correspond to high-inclination and low inclination sources, respectively. For clarity's sake, in each figure we display the data in the following way: in the top panel we show the data for high inclination sources in blue and the data from low-inclination sources in dark grey. In the middle panel we show the data from low inclination sources in red  and the data from high-inclination sources in light grey. In the bottom panel the light and dark grey dots are the same as in the top and middle panes, while the blue circles and red squares are \textquoteleft average\textquoteright\ points, obtained applying a logarithmic rebin in frequency to the grey points. The blues circles and the red squares represent data from high and low-inclination sources, respectively. 

In Fig. \ref{fig:CB_high_low} we compare the distributions of rms for type-C and and type-B QPOs. In Fig. \ref{fig:CBnoise_high_low} we do the same, but this time we consider the noise associated to type-C and type-B QPOs. In fig. \ref{fig:CB_high_low_frequency} we compare the distributions of QPO centroid frequency for type-C and type-B QPOs. 
The plots in Fig. from 3 to 5 show the histograms and the corresponding empirical cumulative distribution functions (ECDF). Blue represents high inclination, red low inclination and an overlap of the two colors in a histogram is indicated by mixing them into purple. As in the previous plots, we  do not separate the different sources.

\subsection{Statistical analysis}\label{sec:statistics}
From Fig. \ref{fig:QPO_vs_freq} it is clear that at least the Type-B and Type-C QPO rms is different in high and low-inclination sources in a quite large frequency range.  To test the significance of these differences we carried out a two-sample Wilcoxon hypothesis test (U-test, also known as Mann-Whitney test, \citealt{Mann1947}).  This is a non-parametric rank sum test designed to check for a difference in location shift. The idea is similar to testing the significance of the difference between two sample means in the popular Student's t-test. However, in our case many of the measured distributions are clearly not normal and therefore the t-test cannot be used. The only assumption of the U-test is that the shape of the two distributions is approximately similar. This is fulfilled here. 

For the distributions in Fig. \ref{fig:CB_high_low}, where the similarity of shape might be doubted, we also performed a Kolmogorov-Smirnov test (KS-test), a different non-parametric test, which does not have this limitation. However, in order to interpret easily the results of the  KS-test, the two ECDF should not intersect, a requirement that is clearly not fulfilled in the majority of cases in Figs. \ref{fig:CB_high_low} to \ref{fig:CB_high_low_frequency}. Therefore, we based our statistical inference primarily on the U-test. This analysis was performed within the R software environment \citep{R_manual}.

In Table \ref{tab:tests} we summarise the results of the U-test for the high- and low-inclination ECDF in Figs. \ref{fig:CB_high_low} to \ref{fig:CB_high_low_frequency}. If the plots suggested an obvious difference between two ECDF we used an one-sided test, i.e. we tested whether there is a location shift into a specific direction (e.g. red is right of blue). For cases in which there was no clear visible difference we tested for a non-zero location shift (i.e. two-sided test). Table \ref{tab:tests} lists, for each test, the resulting p-value, which is the probability of obtaining a difference at least as large as observed if the underlying distributions were in fact identical. 
%
%
\begin{table*}
\renewcommand{\tabcolsep}{0.9cm}
\centering
\caption{QPO, noise and total rms from all the observations on the sources of our sample. We separated type-C and type-B QPOs and high and low inclination sources. The full table is available online only. }\label{tab:parameters}. \\                                                                                                                                                                                                   
\begin{tabular}{c c c c c c }															
\hline                                                                                                                                                                                                                                                                                                                                                                                                                         																		
$\#$            & ID      & Frequency & QPO rms  &  Noise rms  & Total rms   \\                                                                                                                                                                                                                								
                &         & $[Hz]$    & $\%$    &   $\%$          &  $\%$        \\                                                                                             
                						
\hline                                                                                                                                                                                                              \hline                                                                                                                                                                                                                                                                                                                                                                                                                         
\multicolumn{6}{|c|}{ {\bf Type-C} - Low inclination ($< 70^{\circ}$)} \\                                                                                                                                                                                                                      \hline																				\hline
						
\multicolumn{6}{|c|}{Swift	J1753.5--01}	\\																							
\hline																		
														
1	&	91094-01-01-00	&	0.639	$_{-	0.009	}^{+	0.009	}$&	6.7	$_{-	0.6	}^{+	0.6	}$&	33.5	$_{-	0.6	}^{+	0.7	}$&	34.1	$_{-	0.6	}^{+	0.7	}$\\
2	&	91094-01-01-01	&	0.823	$_{-	0.006	}^{+	0.006	}$&	10.5	$_{-	0.6	}^{+	0.6	}$&	30.7	$_{-	1.1	}^{+	1.2	}$&	32.5	$_{-	1.0	}^{+	1.1	}$\\
3	&	91094-01-01-02	&	0.842	$_{-	0.005	}^{+	0.005	}$&	9.6	$_{-	0.5	}^{+	0.5	}$&	31.9	$_{-	1.5	}^{+	1.6	}$&	33.3	$_{-	1.4	}^{+	1.5	}$\\
4	&	91094-01-01-03	&	0.824	$_{-	0.005	}^{+	0.005	}$&	12.3	$_{-	0.5	}^{+	0.5	}$&	30.2	$_{-	3.3	}^{+	3.8	}$&	32.6	$_{-	3.1	}^{+	3.6	}$\\
5	&	91094-01-01-04	&	0.832	$_{-	0.006	}^{+	0.006	}$&	14.6	$_{-	0.7	}^{+	0.8	}$&	29.0	$_{-	1.3	}^{+	1.4	}$&	32.5	$_{-	1.2	}^{+	1.3	}$\\
... \\

\hline
\end{tabular}
\end{table*}
%
\subsection{Results from the tests}\label{sec: results_from_tests}
%
\begin{figure}
\centering 
\includegraphics[width=8.4cm]{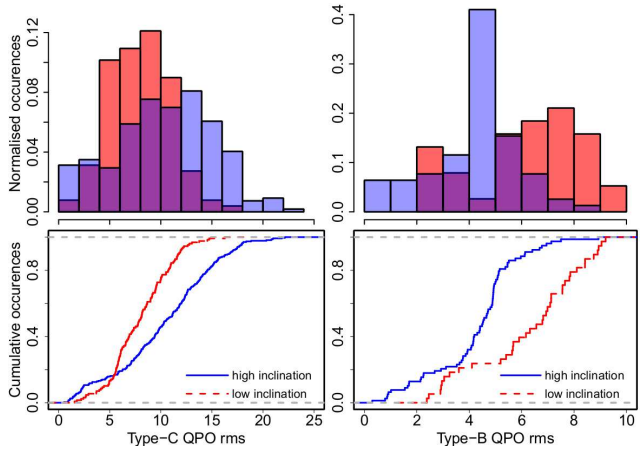}
\caption{Type-B and type-C QPO rms distributions and correspondent ECDF. The distributions are different with high significance. Histograms are normalized to unit area. A color version of the figure is available online.}\label{fig:CB_high_low}
\end{figure}
\begin{figure}
\centering
\includegraphics[width=8.4cm]{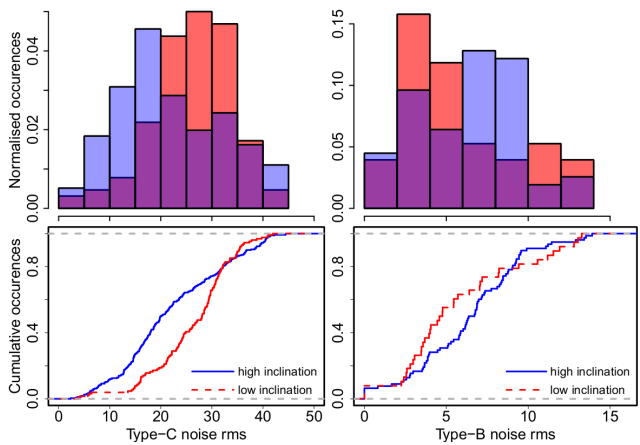}
\caption{Type-B and type-C QPO associated noise rms distributions and correspondent ECDF. There is a  significant difference for the type-C QPO noise distributions of about 4$\sigma$-equivalent confidence level. Histograms are normalized to unit area. A color version of the figure is available online.}\label{fig:CBnoise_high_low}
\end{figure}
\begin{figure}
\centering
\includegraphics[width=8.4cm]{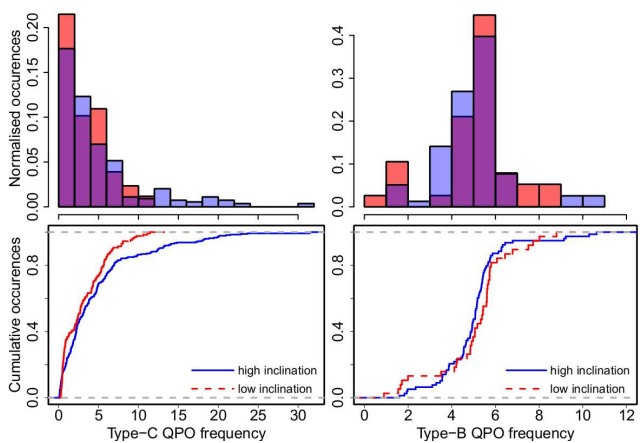}
\caption{Type-B and type-C QPO frequency distributions for high- and low-inclination sources.  The high- and low-inclination distribution for both type-C and type-B QPOs are broadly comparable. Histograms are normalized to unit area. A color version of the figure is available online.}\label{fig:CB_high_low_frequency}
\end{figure}
%
From the statistical tests described in Sec. \ref{sec:statistics}, we find that the rms distributions of both type-C and type-B QPOs  of high-inclination vs low-inclination sources are significantly different. 
%
\begin{itemize}

\item Figure \ref{fig:CB_high_low}: here we compare the QPO rms distribution for type-C and type-B QPOs. For both QPOs the distributions of high and low-inclination sources are very different. The type-C QPO rms appears to be, on average, higher for high-inclination sources, while the opposite is true for type-B QPOs that are stronger in low-inclination systems. In the case of type-C QPOs, the high and low-inclination distributions are different mainly at intermediate frequencies ($\sim$0.5--10 Hz), while they overlap at lower and higher frequencies (see Fig. \ref{fig:QPO_vs_freq}). Also in the case of type-B QPOs the distributions are significantly different and both a U and a KS-test\footnote{The KS-test is sensitive to location shifts as well as differences in the distribution shape} give similar results. 

\item Figure \ref{fig:CBnoise_high_low}: this plot compares the rms of the noise associated with type-C and type-B QPOs. Again we separate type-B and -C QPOs. While the high- and low-inclination populations of type-B QPOs noise rms distributions are not significantly different, we detect a significant difference (about 4$\sigma$) in the case of type-C QPO noise, with the noise stronger for low-inclination sources (i.e. the noise does the opposite with respect to the QPOs). However, the difference is mainly due to the points between 1 and 8Hz (see Fig. \ref{fig:noise_vs_freq}).

\item Figure \ref{fig:CB_high_low_frequency}: this plot compares the QPO frequencies for type-C and type-B QPOs separately. The high and low-inclination type-B QPO frequency distributions are broadly comparable, while the two type-C QPO frequency distributions are marginally different. This is probably due to the high-frequency tail in the high-inclination type-C QPO frequency distribution. This feature decreases the similarity between the distributions and the formal test should therefore be interpreted with care.

\item For completeness, we also compared the total rms distributions from the observations where a type-C or a type-B QPO was found. The distributions do not show significant differences. In the case of type-C QPOs, this is probably due to the fact that the QPO and the associated noise distributions show an opposite dependence on inclination that shifts the total rms distributions closer together.
In the case of Type-B QPOs the total rms distribution is dominated by the noise (inclination independent) and the small difference between the type-B distributions is probably due to the presence of the (significantly different) QPO rms. 

\end{itemize}

\begin{figure*}
\centering
\includegraphics[width=18cm]{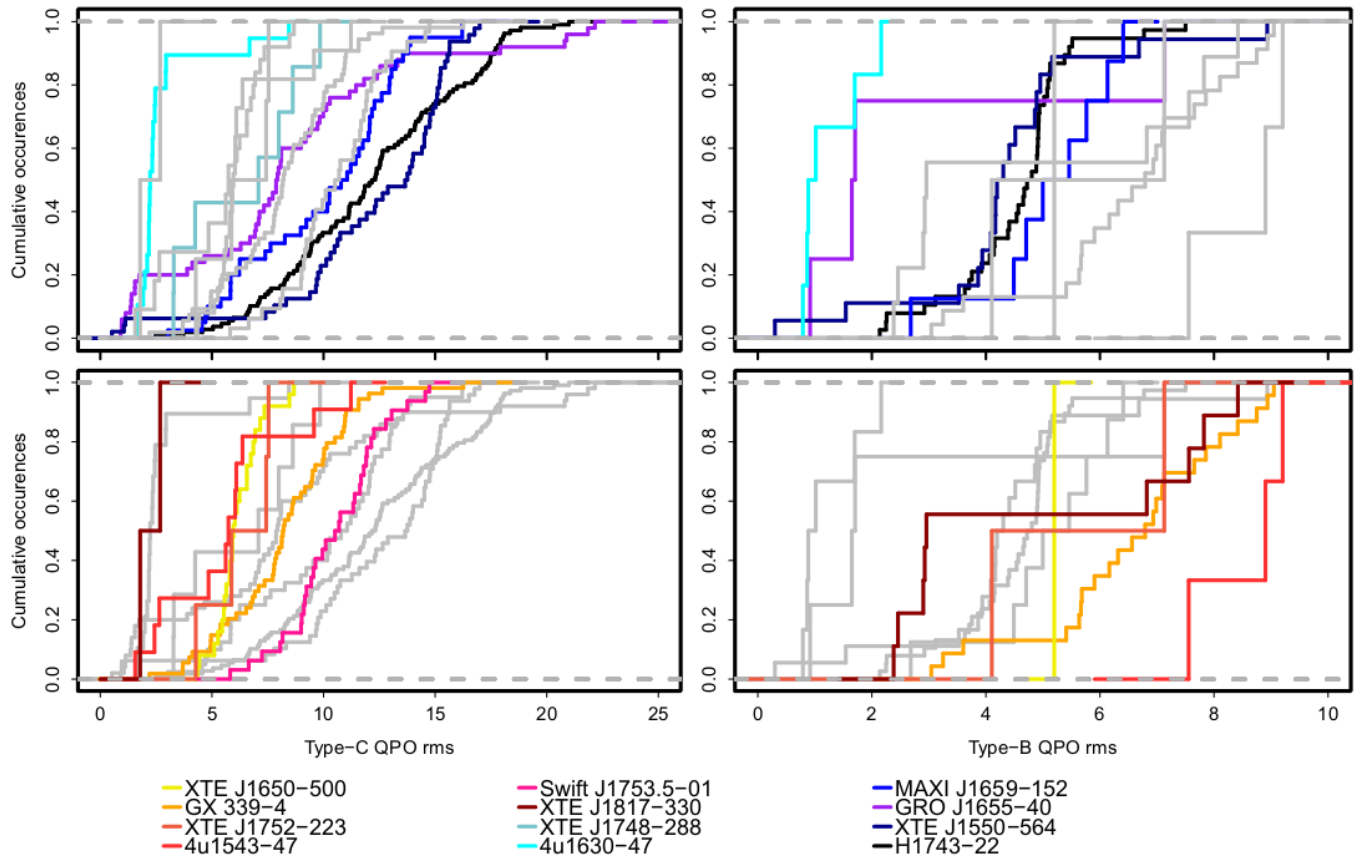}
\caption{Cumulative QPO rms distributions (for type-C and B QPOs, left and righ panels, respectively)  for the individual sources of our sample. Different colors correspond to diffferent sources (see legend). To facilitate the comparison between the two population of sources, we first show the high-inclination sources in different colors and the low inclination sources in grey (top panels) and then the low inclination sources in colours and the high inclination sources in grey.  A color version of the figure is available online.}\label{fig:C_rms_inclination}
\end{figure*}

In order to study in greater detail the dependence of the amplitude (the rms) of the QPO on inclination, we produced two additional plots. 
The top and bottom left panels in Fig. \ref{fig:C_rms_inclination} combine the ECDFs of the type-C QPO rms for all the individual sources. The right panels show the same for type-B QPOs. High-inclination sources (top panels) are plotted  in cyan, light blue, blue, purple, dark blue and black and low-inclination sources (bottom panels) are represented in yellow, orange, dark orange, red, pink and  brown. In the case of type-C QPOs there is some overlap, but high and low-inclination sources can clearly be distinguished. The situation is somewhat less clear for the type-B QPO rms distributions, because of the dominance of GX 339-4, and maybe the fact that a couple of sources have so few measurements. XTE J1650-500 and XTE J1752-223 only displayed one and two type-B QPOs, respectively, therefore they do not provide any valid information. 

Since GX 339--4 provides more type-B QPOs to the low-inclination group than any other source, we performed the tests described in Sec. \ref{sec:statistics} also excluding this source from the sample. For both the type-C and type-B noise distributions, as well as in the type-C QPOs distributions, we obtain similar results to what we report in Tab. \ref{tab:tests} (low vs high inclination average QPO rms distributions without GX 339-4:  p-value for type-C QPO is 2.033e-06). However, in the case of type-B QPOs we found that the difference in the high and low-inclination distributions becomes only marginally significant (low vs high inclination average QPO rms distributions without GX 339-4: p-value for type-B QPO is 0.023), which could be mainly due to the reduction in sample size (number of data points reduced from 38 to 15). Thus, it appears that the contribution from GX 339-4 dominates the low-inclination type-B QPO population. We also tested for significant differences between the QPO rms ECDF of GX 339-4 and the average low-inclination QPO rms distributions and we found no significant differences (GX 339-4 vs low-inclination average QPO rms distributions: p-value for type-C QPO is 0.397; p-value for type-B QPO is 0.46), suggesting that GX 339-4 shows no obvious deviation from the average low-inclination sources behavior. Therefore, our results still hold despite the dominance of GX 339-4. 

%
\begin{table}
\centering
\caption{Results of the statistical tests performed on the parameter distributions for the type-C, type-B QPOs and their associated noise. \textquoteleft a\textquoteright\ and \textquoteleft b\textquoteright\ indicate that a one-sided or a two-sided test, respectively, has been performed. In the \textquoteleft significance\textquoteright\ column we indicate whether the distributions we compare in each line are significantly different. S: distributions are significantly different (p-value $<$ 0.01), MS: distributions are marginally different (0.01 $\leq$ p-value $\leq$ 0.05), NS: distributions are not significantly different (p-value $>$ 0.05). 
}
\label{tab:tests}
\begin{tabular}{c c c }
\hline
				low vs high inclination					&		p-value  & Significance \\
\hline
\hline

Type-C \textbf{QPOs} rms 	& 3.421$\times 10^{-8}$	 (U-test) $^a$ 	&  S 		\\
         					& 8.482$\times 10^{-9}$	 (KS-test)	&  S 		\\

Type-B \textbf{QPOs} rms 	& 1.047$\times 10^{-6}$	(U-test) $^a$	&  S		\\
							& 5.857$\times 10^{-8}$	(KS-test) 	&  S 		\\
\hline 

Type-C QPOs \textbf{noise} rms 	& 1.108$\times 10^{-5}$ (U-test) $^b$	&  S		\\

Type-B QPOs \textbf{noise} rms 	& 0.1686	 (U-test) $^b$				&  NS 		\\

\hline

Type-C QPOs frequency 			& 0.044	 (U-test) $^b$				&  MS 		\\

Type-B QPOs frequency 			& 0.169 (U-test) $^b$				&  NS 		\\

\hline
\end{tabular}
\end{table}
\subsection{The case of XTE J1859+226 and MAXI J1543-564}\label{sec:1859}

\begin{figure}
\centering
\includegraphics[width=8.25cm]{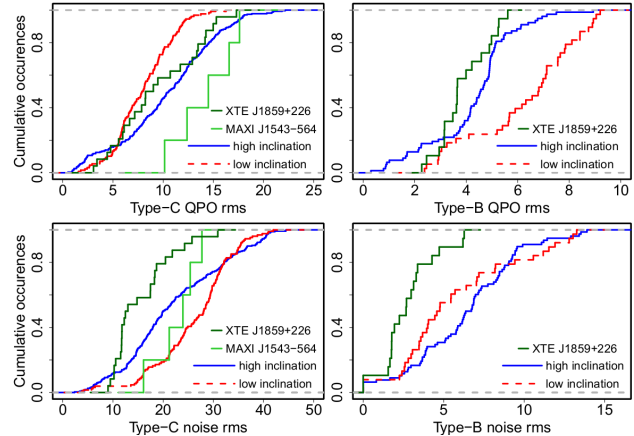}
\caption{Cumulative QPO rms distributions (for type-C and B QPOs, top left and top righ panels, respectively) and noise rms distributions (associated with type-C and -B QPOs, bottom left and bottom right panels, respectively) for XTE J1859+226 (dark green line) and MAXI J1543-564 (ligh green line). The red and blue line correspond to the average ECDF of low and high-inclination sources, respectively, as it is shown in Fig. \ref{fig:CB_high_low}. Note that in the case of Type-B QPOs, only XTE J1859+226 is shown since MAXI J1543-564 did not show any significant type-B QPOs during its outburst. A color version of the figure is available online.} \label{fig:intermediate_sources}
\end{figure}

As noted in Sec. \ref{sec:sample}, we could not unambiguously classify XTE J1859+226 and MAXI J1543-564 as high or low-inclination systems. Therefore, we treat them separately here. 
In Fig. \ref{fig:intermediate_sources} we show a comparison between the QPO and noise rms ECDF for XTE J1859+226 and MAXI J1543-564 and the correspondent average ECDF for high and low-inclination systems (shown in Fig. \ref{fig:CB_high_low}). We see that in the relevant cases (Type-B/C QPO rms and type-C noise), despite some overlap with the average distributions, both systems qualitatively resemble the behavior of high-inclination sources rather than low-inclination sources. We note, however, that we only have 5 type-C QPOs from MAXI J1543--564, therefore the statistics is fairly low in this case. 
\section{Discussion}\label{sec:discussion}

We used a large archival data-set from the RXTE satellite to measure, with a \textquoteleft population-statistics approach\textquoteright, the amplitude of the aperiodic variability (i.e. QPO amplitude and noise amplitude) in a number of BH X-ray transients. 
We collected a total of 564 QPOs: 128 type-C QPOs and 38 type-B QPOs from six low-inclination sources and 272 type-C QPOs and 78 type-B QPOs from six high inclination sources. In addition, we collected 29 type-C QPOs and 19 type-B QPOs from intermediate-inclination sources.

These are our main findings:
\begin{itemize}

\item We find a highly significant difference between the QPO rms distributions of  high-inclination vs low-inclination sources. This is true for both type-C and type-B QPOs (even though the low-inclination type-B QPO population is somewhat dominated by GX 339-4). Our study confirms results obtained for a smaller sample of sources and observations by \cite{Schnittman2006} for the case of type-C QPOs. 
The QPO amplitude correlates significantly with the inclination of the binary system: type-C QPOs are generally stronger in high inclination sources (i.e. closer to edge on), while type-B QPOs are generally stronger in low-inclination sources (i.e. closer to face-on sources). 

\item We find a significant difference in the noise distributions associated with type-C QPOs, while in the case of type-B QPOs the associated noise seems to be inclination independent. On average, the noise is consistent with being stronger for low-inclination sources (at least in a certain frequency range), at variance with the behavior of QPOs. This constitute the first evidence that the broad band noise associated with the QPOs shows a different behavior with respect to the QPO themselves and, for this reason, could have a separate origin.


\item Comparing these results with the properties of XTE J1859+226 and MAXI J1543-564, we conclude that the inclination of these sources is consistent with being intermediate to high rather than low inclination, in line with what the properties listed in Tab. \ref{tab:sources} suggested. Of course, accurate inclination measurements for both sources are needed to confirm this result. 
\end{itemize}
Recently, with a grouping very similar to what we used, but through a different method, \cite{Heil2014} found the same inclination-dependence in the amplitude of type-C QPOs that we have found. \cite{Heil2014} also reported  that the broad band noise associated with type-C QPOs is inclination independent. However, since the comparison between high and low-inclination sources performed by these authors is qualitative, their results are still broadly consistent with ours.\\

\subsection{Caveats}

\subsubsection{The sample} 

Our sample of sources is quite small and tests on a different, independent sample of sources are needed to confirm our findings. Furthermore, in this work we focused only on the general differences in the QPOs and noise distributions, without investigating the details of the distributions themselves. 
However, the different dependence of the QPO rms on the orbital inclination already suggest that type-B and type-C QPOs are intrinsically different and could come from two well-defined geometrical regions. Most likely, they are the effect of different physical processes and/or the result of different geometrical configuration/physical properties (e.g. change in plasma temperature, ionization state, viscosity or optical depth) of the accreting material in the region where they are produced. This is supported by the results obtained by \cite{Motta2012} who discovered simultaneous type-C and type-B QPO in the data of the black hole binary GRO J1655-40. This result essentially ruled out the possibility that type-B QPOs could arise from the same physical phenomenon originating type-C QPOs - regardless of what this phenomenon is - supporting what was already suggested by \cite{Motta2011a} based on the dependence of the different kind of QPOs on the hard emission. 

\subsubsection{Source classification and grouping}
Since the inclination measurements are often not precise and affected by large uncertainties, our source classification mostly relies on the assumption that absorption dips are a strong indication of high orbital inclination. Even though this association is commonly adopted (see Appendix  \ref{App:AppendixB}), it represents a critical caveat for our results, as any new observation showing absorption dips might turn a low-inclination source into an high-inclination one (while it practically impossible for a high-inclination source to turn into a low-inclination one). 
Our choice of considering only well-monitored sources is intended to minimize the possible missing of an high-inclination feature. Even so, the misclassification of one or more sources is still possible as it is known that absorption dips can disappear during an outburst (see e.g. \citealt{Smale1999}, \citealt{Kuulkers2013}) - making their detection harder - or can be detected only after being well-sampled by years of X-ray observations, as it happened in the case of H1743-322 (\citealt{Homan2005b}; see also \citealt{DA`i2014}). 
In order to investigate the effects of such misclassification, we performed again the U-tests described in Sec. \ref{sec:statistics} on the high and low-inclination distributions, but this time moving, one by one, the sources of the low-inclination group to the high-inclination group. 
We see that, predictably, the significance of the difference between the high and low-inclination distributions tends to decrease when moving a low-inclination source to the high-inclination group. However, the difference remains always significant (p-value always $<$ 0.001), with the exception of GX 339-4 in the case of the type-B QPOs rms distributions (p-value = 0.082). This is probably mainly due to the fact that GX 339-4, as we already discussed in Sec. \ref{sec: results_from_tests}, dominates (without biasing it) the low-inclination type-B QPOs rms distribution. We note, however, that GX 339-4\footnote{GX 339-4 has been observed with RXTE 1426 times over a time span of about 15 years.} is the best monitored BHB of our sample and therefore it is also the least likely source to be misclassified.

Additionally, 
we see a considerable scatter within the QPO rms distributions of the high- and low-inclination groups (see Fig. \ref{fig:CB_high_low}). While the average difference between the two groups is highly significant, there is also an overlap of the individual source ECDFs. However, in Fig.\ref{fig:CB_high_low} we clearly see that there are strong differences between the individual sources and that the QPO rms distributions cannot be considered similar. 
Evidently, it is in principle possible to group the sources differently from our categorization based on inclination.  To test the goodness of our grouping, we randomly divided the twelve sources  of our sample  into two groups of six and used the same statistical test as above to compute the significance of the difference between the QPO rms distributions. For 10$^6$ of these random samplings we found that in about 5\% of all cases the result is at least as significant as the one reported in Tab. \ref{tab:tests} (after correcting for the large number of comparisons made). This is in agreement with the scatter visible in Fig.\ref{fig:CB_high_low}.
Nonetheless, unlike all those random groupings, our initial classification was made \textit{a priori} based on physical arguments. Therefore, its statistical significance conveys a physically meaningful result.

\subsubsection{Inner-outer disk misalignment} 
It has been recently suggested (see \citealt{Veledina2013}) that a misalignment between orbital spin and BH spin (presumably aligned with the jet axis) might significantly influence the rms amplitude of type-C QPOs in BHBs. In this context, the relevant angle  for the precession is the angle between the BH spin and the line of sight, therefore a misalignment between spin and orbital plane (not considered here, as we use the angle between the line of sight and the orbital spin) could in principle largely affect the QPO amplitude. It is also worth noticing that the outer disk inclination and the inner disk inclination might differ from one another as the inner disk can be strongly affected by the BH spin, i.e. the inner disk might tend to align with the plane normal to the BH spin, while the outer disk tends to stay aligned with the orbital plane.
The difficulty in estimating the BH spin/orbital spin misalignment constitutes an additional source of uncertainty and, together with the fact that most inclination measurements are poorly constrained,  (especially in the case of nearly face-on sources, see Appendix \ref{App:AppendixB}) prevents us from studying the details of the effects of inclination on the ECDFs. \\



\smallskip
 
\subsection{About the origin of LFQPOs}
 
The existing models that attempt to explain the origin of LFQPOs are generally based on two different mechanisms: instabilities and  geometrical effects. In the latter case, the physical process typically invoked is precession. \\
\cite{Titarchuk2004} proposed the so called \textit{transition layer model}, where type-C QPOs are the result of viscous magneto-acoustic oscillations of a spherical bounded transition layer, formed by matter from the accretion disc adjusting to the sub-keplerian boundary conditions near the central compact object. This mechanism would not produce inclination dependent QPOs and is, therefore, unable to explain our results.\\
\cite{Cabanac2010} proposed a model to explain simultaneously type-C QPOs and the associated broad band noise. Magneto-acoustic waves propagating within the corona makes it oscillate, causing a modulation in the efficiency of the Comptonization process on the embedded photons. This should produce both the type-C QPOs (thanks to a resonance effect) and the noise that comes with them. This model predicts that both type-C QPOs and their associated noise would be inclination-independent. Therefore, also the predictions of this model do not match the observed properties of the LFQPOs reported here. \\
\citealt{Tagger1999} proposed a model based on the \textit{accretion ejection instability} (AEI), according to which a spiral density wave in the disc, driven by magnetic stresses, becomes unstable by exchanging angular momentum with a Rossby vortex. 
This instability forms low azimuthal wavenumbers, standing spiral patterns which would be the origin of LFQPOs. 
\cite{Varni`ere2002} and \cite{Varni`ere2012} suggested that all the tree types of QPOs (A, B and C) can be produced through the AEI in three different regimes: non-relativistic (type-C), relativistic (type-A, where the AEI coexist with the Rossby Wave Instability (RWI), see \citealt{Tagger1999}) and during the transition between the two regimes. \cite{Varni`ere2005} investigated the impact of a clean spiral density wave on the emission from an accreting BH and found that the rms of type-C QPOs is expected to be higher when the disk is close to edge on  (\citealt{Varni`ere2005}), therefore the AEI model succeeds in partly reproduce our findings. However, as the effects of the interplay of the RWI and the AEI must be taken into account, further investigation is needed to fully understand the predicted behavior of type-B and type-A QPOs, as well of their associated noise (Varni\`ere, private communication). 

In summary, all the models based on instabilities fail in fully explaining the properties of the QPOs and of the noise. Therefore, we are left only with the models involving geometrical effects. 
The maximum radius at which a $\sim$Hz modulation can be produced in a 10$M_{\odot}$ BH would come from orbital motion at $>$100 R$_g$ from the BH. Hence, in principle, type-C QPOs could be the result of modulation/occasional obscuration of the emission from the inner hot flow operated by structures in the disk (e.g. clumps in the accretion flow). However, in this case, a mechanism producing both obscuring structures - that would need to be unrealistically vertically extended to be able to partially obscure the emission from the inner flow - and a mainly equatorial emission from the inner flow is necessary to explain what we observed. Furthermore, \cite{Motta2014} and \cite{Motta2014a} have shown that the Lense-Thirring mechanism successfully explain the evolution of type-C QPOs (and other timing features) in two BHBs (GRO J1655-40 and XTE J1550-564).

\cite{Ingram2009} proposed a model based on the relativistic precession as predicted by GR that attempts to explain type-C QPOs and their associated noise. 
This model requires a cool optically thick, geometrically thin accretion disc  (\citealt{Shakura1973}) truncated at some radius, filled by a hot, geometrically thick accretion flow. This geometry is known as \textit{truncated disc model} (\citealt{Esin1997}, \citealt{Poutanen1997}). 
In this framework, type-C QPOs arise from the Lense-Thirring precession of a radially extended section of the hot inner flow that modulates the X-ray flux through a combination of self-occultation, projected area and relativistic effects that become stronger with inclination (see \citealt{Ingram2009}).  
The broad-band noise associated with type-C QPOs, instead, would arise from variations in mass accretion rate from the outer regions of the accretion flow that propagate towards the central compact object, modulating the variations from the inner regions and, consequently, modulating also the radiation in an inclination-independent manner (see \citealt{Ingram2013}). The predictions of this model matches our findings if the anisotropy of the hot-flow emission is considered. \cite{Veledina2013} simulated the X-ray and optical emission predicted by this model for variable source inclination angles, taking into account also the possible relativistic effects on the accretion flow. From Fig. 3 in \cite{Veledina2013}, we infer that the angular dependence of the coronal emission varies with the distance from the BH. In particular, X-ray emission from the outer regions of the hot-flow (tens to hundreds gravitational radii) is more visible in low-inclination sources, while the opposite is true for the inner regions (3-5 gravitational radii), whose emission is more visible in high-inclination sources. Therefore, low-inclination sources should have a softer coronal emission than high-inclination sources. The broad band noise is also known to have a soft spectrum, at least in the hard states (\citealt{Gierli'nski2005}). Although little can be said about the intrinsic energy dependence of the noise in the intermediate states (because of the soft emission from the disk) it is reasonable to assume that the noise itself does not change much its properties from the hard to the intermediate state, thus remaining soft (i.e. coming predominantly from the outer regions of the hot flow). This implies that the noise associated to type-C QPOs is expected to be stronger in face on sources, as we do observe in our data (at least when the QPO is found in a certain frequency range). In conclusion, the predictions of the model  proposed by \cite{Ingram2009,Ingram2011,Ingram2012,Ingram2013}  are consistent with the observed properties of both type-C QPOs and noise.

\smallskip

\noindent Type-B QPOs show an opposite behaviour to type-Cs, i.e. their amplitude is larger for low-inclination sources. Only the model by \cite{Tagger1999} attempts to explain the origin of type-B QPOs, however our findings points to an intrinsic difference between type-C and type-B QPOs, which are still not fully explained by their model. 

Type-B QPOs have been tentatively associated with the relativistic ejections usually observed along the hard-to-soft transitions in transient BHBs (\citealt{Fender2004}), even though a clear cause-effect relation between the two phenomena  has not been identified (\citealt{Fender2009}, \citealt{Miller-Jones2012}). 
There are two main ways in which the observed dependence on inclination of type-B QPOs could be produced, assuming that the type-B QPOs come from the jet: (i) a lack of obscuration when the observer is able to look directly into the jet; (ii) relativistic aberration due to the jet flow. In case (i) the oscillator responsible for the production of type-B QPOs needs not be physically associated with the relativistic outflow, while in case (ii) it would actually be part of the jet. However, in case (i) we would expect abrupt changes in the amplitude of the type-B QPO (see \citealt{Nespoli2003}), while more continuous changes would be expected in case (ii). In any case, the currently available data and the poor inclination constraints do not allow to confirm either hypothesis. 

Despite the fact that the launch of the relativistic jet are expected to provoke significant changes in the structure of the accretion flow, only very small spectral changes are observed in the X-ray spectra in correspondence to the radio flares associated with the relativistic jets (a few \% in the hard emission, see \citealt{Motta2011}). The sole dramatic changes observed (almost) simultaneously to the launch of the jets are the  transitions between type-C and type-B QPOs (see e.g. \citealt{Miyamoto1991a}, \citealt{Takizawa1997}). We note, however, that it has been suggested that major radio flares seen across the hard to soft transition could be produced by internal shocks associated with varying/increasing jet velocities. If this is the case, only small changes are expected in the X-ray energy spectra and PDS (see e.g. \citealt{Malzac2013}). 

Our results support the hypothesis that type-B QPOs are related to the relativistic jet (\citealt{Fender2009}), since there is no other obvious mechanism that would be stronger face-on. Therefore, our results support what has been already proposed by \citealt{Fender2004} and \citealt{Fender2009}: the transitions between type-C and type-B QPOs track the dramatic changes in the configuration and/or in the physical properties of the accretion flow linked to the launch of relativistic jets (but see the case of GX 339-4, 2002 outburst, in \citealt{Fender2009}). As the mechanism of production of type-B QPOs and the launching mechanism of the jets is still unknown, further evidences are required to confirm this statement. 
Finally, the fact that the noise associated with Type-B QPOs seems to be inclination independent might indicate that the type-B QPOs noise is produced by yet another mechanism, that is not sensitive to inclination effects. 

\section{Summary and conclusions}								
						
We have analyzed a large sample of archival RXTE observations were we detected low frequency QPOs. 
We assumed that there are no intrinsic differences between the sources of our sample and that the presence of absorption dips in the light-curve of a source corresponds to high orbital inclination. 
We have shown that inclination has a strong effect on the QPOs. We found that:

\begin{itemize}
\item Type-C QPOs appear stronger in high inclination sources.
\item Type-B QPOs  show the opposite behavior, being stronger for low-inclination sources. 
\item The noise associated with both type-C QPOs is consistent with being stronger for low-inclination sources, while the noise associated with type-B QPOs is  consistent with being inclination independent. 
\end{itemize}

Our results suggest that:

\begin{itemize}

\item type-C QPOs, type-B QPOs and the broad band noise associated with type-C QPOs are geometrically/physically different phenomena.

\item type-C QPOs are consistent with having a geometrical origin. In particular, we find that the relativistic precession is the only mechanism that satisfies all our observational constraints and therefore is favoured by our results.

\item at variance with type-C QPOs, the associated broad band noise might, instead, correspond to intrinsic brightness variability induced by fluctuations in the mass accretion rate propagating in an hot flow that emits in a non-isotropic way.

\item fast transition between type-C and type-B QPOs could be the best trackers in the X-rays of the relativistic ejections typical of  most BH transients. 

\end{itemize}

%
%
%

\bigskip
\noindent
\footnotesize{SEM acknowledges Peggy Varniere, Lucy Heil, Erik Kuulkers and Jari Kajava for useful comments and discussion on this work. SEM and MH acknowledge support from the ESA research fellowship program and from the ESAC Faculty. SEM also acknowledges the Observatory of Rome and Brera for hospitality. TMD and PC acknowledge support from the ESAC Faculty and ESA for hospitality.
TMB and SEM acknowledge support from INAF PRIN 2012-6. TMD acknowledges funding via an EU Marie Curie Intra-European Fellowship under contract no. 2011-301355. PC acknowledges support by a Marie Curie FP7-Reintegration-Grants under contract no. 2012-322259. 
This research has made use of data obtained from the High Energy Astrophysics Science Archive Research Center (HEASARC), provided by NASA's Goddard Space Flight Center. 
}
%
%
%
%
%

\appendix
\section{On the relation between high-inclination and absorption dips} \label{App:AppendixA}

Absorption dips in the X-ray light-curves usually recur at the orbital period of a system and are thought to be caused by obscuration by material located in a thickened outer region (`bulge') of the accretion disk. The bulge forms as a consequence of the interaction of the accretion disk with the inflowing gas stream from the companion star (see e.g., \citealt{White1982},\citealt{Walter1982}). The presence of absorption dips is commonly considered a signature of high orbital inclination (but see \citealt{Galloway2012}\footnote{\citealt{Galloway2012} reported apparent dipping activity early in the 2011 outburst of the NS X-ray binary Aql X-1. Such features had not been previously reported in Aql X-1, but resembled the absorption dips observed in other X-ray binaries. \cite{Galloway2012} therefore concluded that, since apparent dipping behaviour can occur at times, the system inclination is at the high end of its likely range (36$^{\circ}$-55$^{\circ}$), \citealt{Robinson2001}}) and can allow measurement of the orbital period of a source (\citealt{White1982}, \citealt{White1985}, \citealt{DiazTrigo2009}, \citealt{Casares2014}). 
Absorption dips in low-mass X-ray binaries (LMXBs) have been observed mainly around orbital phase 0.7--0.9, while eclipses are expected at phase zero when the accretion disk is seen at $\sim$90$^{\circ}$, i.e., when the companion star is closest to us and in front of the compact object (e.g., Parmar \&\ White 1988). Occasionally, `secondary absorption dips'  are observed at a 0.5 phase difference with respect to the phase at which `regular dips' occur are also observed. They are explained as being due to material 
migrating to the other side after the impact with the disk (see, e.g., \citealt{Frank1987}, \citealt{Armitage1998}, and references therein). 

The orbital inclination angle with respect to the line of sight necessary to produce absorption dips depends both on the geometry of the system and on the properties of the accretion flow (e.g. the size of the accretion disk and of the bulge) and therefore might be different for every source. However, it is still possible to estimate the minimum inclination angle that would provoke absorption dips in a standard BHB. 
This lower limit on the inclination is based on the size of the bulge and not on the disk opening angle, which has been generally estimated to be $\simeq$12$^{\circ}$ (e.g., \citealt{deJong1996}, \citealt{Bayless2010}). \cite{White1982a} estimated the size of the bulge responsible for absorption dips as 19$^{\circ}\pm$6$^{\circ}$ for the LMXB 4U\,1822$-$37. 
Taking this value as typical for LMXBs, we can thus set a lower limit on the disk inclination of 65$^{\circ}$. We note, however, that if the accretion disk is tilted or warped, the lower limit
for the inclination could be as low as 55$^{\circ}$, by considering that for a generic LMXBs, a disk tilt of about 10$^{\circ}$ is expected (\citealt{Foulkes2010}).

In several LMXBs the absorption dips appear for only part of the outburst (see, e.g., in MAXI\,J1659$-$152, \citealt{Kuulkers2013}, 4U\,1630$-$47 and GRO\,J1655$-$40, \citealt{Kuulkers1998}, \citealt{Kuulkers2000}, \citealt{Tomsick1998}, H1743-322, \citealt{Homan2005}). The explanation for this lies on the fact that for transient BHXBs, changes in the accretion mode cause the appearance or disappearance of dips. For instance, \cite{Kuulkers2000} interpreted the (deep) absorption dips during the rise and plateau phase of the outburst in GRO J1655$-$40 as due to filaments in the stream of material coming from the companion star and splashing into the accretion disk, overflowing above and below the impact region. If the inclination is high enough, the impact region itself comes also into the line of sight (e.g., \citealt{Frank1987}). However, the presence of absorption features all around the orbit for neutron stars (e.g., \citealt{Parmar2002}) shows that at least part of the 
photo-ionised plasma is distributed equatorially along the whole plane of the disk, indicating that absorption is due to a structure in the disk rather than by filaments. In that scenario, the cause for the disappearance of dips in BHXBs could be, e.g., a strong ionisation of the plasma in bright (but hard) states of the outburst, which renders the plasma transparent and therefore invisible to the observer. An alternative explanation could be that a change of the structure of the accretion flow could diminish the thickness of the bulge and cause the absorption dips to disappear.

\section{On the inclination of the sources of our sample} \label{App:AppendixB}
The inclination angle of a binary with respect to the line of sight can be obtained, in absence of eclipses, through indirect methods based on information from the lightcurve and the spectrum of the optical companion star. The binary inclination is commonly obtained through fitting optical/NIR light curves with synthetic ellipsoidal models, since the amplitude of the modulation normally observed in the OIR lightcurves strongly depends on the inclination angle. However, the vast majority of transient XRBs posses a faint K-M donor star and therefore their lightcurves can be seriously contaminated by other non-stellar sources of light (e.g. the outer accretion disk), affecting critically the inclination measurement.

\subsection{Swift J1753.5-0127}

SWIFT J1753.5−0127 is an X-ray transient discovered by the Swift/BAT on 2005 as a bright variable X-ray source (\citealt{Palmer2005}). Although the mass of the primary has not been dynamically measured yet, the system displayed a number of characteristics that suggest that the binary hosts a BH (\citealt{Soleri2013}).

\cite{Neustroev2014} reported results from optical and UV observations of Swift J1753.5-01. Despite the fact that very low values for inclination are to be excluded (e.g.  no absorption lines - typical of high-inclination bright X-ray novae - and orbital-like modulation in the light-curve that suggest inclination larger than 40$^{\circ}$), high orbital inclination values must be also excluded for two reasons. First, extensive photometry and spectroscopy of Swift J1753.5-01 rule out any significant absorption dip or eclipse, commonly considered a signature of high-inclination. Second, given the orbital parameters of Swift  J1753.5-01, inclinations larger than $\sim$55$^{\circ}$ are highly improbable, since it would require a physically unacceptable black-hole mass smaller than $\sim$2M$_{\odot}$.

\subsection{4U 1543--47}

4U 1543--47 is a bright soft X-ray transient, firstly observed in outburts in 1971 (\citealt{Matilsky1972}). Based on its spectral properties, the compact object has been classified as a BH candidate. 

The presence of (small) ellipsoidal modulations in the optical light curve of the source allowed \cite{Orosz1998} to constrain the inclination between 24$^{\circ}$ and 36$^{\circ}$ (obtained fitting the V and I optical light-curves), with extreme hard lower and upper limits at 20$^{\circ}$ and 40$^{\circ}$, respectively. The second set of limits takes into account possible systematics that the ellipsoidal modelling does not account for.

\subsection{XTE J1650--500}

XTE J1650-500 was discovered by RXTE in 2001 (\citealt{Remillard2001}) and classified as a strong BH candidate based on its X-ray spectrum and variability properties (see e.g. \citealt{Wijnands2001}). 

\cite{Orosz2004} performed a photometric and spectroscopic analysis of XTE J1650-500, and reported a lower limit to the orbital inclination of $\sim$47$^{\circ}$, assuming no disk contamination of the optical light-curve. Because of lack of eclipses, \cite{Orosz2004} concluded that the inclination must be lower than $\sim$70$^{\circ}$, even though the exact value depends on the mass ration on the binary. However, these authors noted that to obtain an inclination of $\sim$70$^{\circ}$  the accretion disk should contribute 80\% of the optical emission. Additionally, \cite{Orosz2004} showed that an inclination higher than 70$^{\circ}$ would yield a BH mass lower than 4M$_{\odot}$.

\subsection{GX 339--4}

GX 339-4  was discovered by OSO 7 in 1972 (\citealt{Markert1973}).  Even when GX 339-4 was first discovered, its aperiodic X-ray variability on timescales from milliseconds to seconds suggested that the compact object in the system is a BH. This transient source is among the most studied by RXTE, having shown 4 outbursts in the last decade. 

\citealt{Cowley2002} reported the results from simultaneous optical photometric and spectroscopic observations. Only small amplitudes in the emission-line
velocity were visible, suggesting that GX 339-4 is seen at a low orbital inclination angle. This is confirmed by the lack of eclipses and absorption dips in both the optical and X-ray light-curves (\citealt{Munoz-Darias2008}), that allow to set a upper limit on the inclination of $\sim$70$^{\circ}$. However, a lower limit to the inclination can be set assuming that the BH mass should not exceed 20M$_{\odot}$: under this hypothesis, the large estimate of the mass function for GX 339-4 implies an inclination larger than 40$^{\circ}$.

\subsection{XTE J1752--223}

XTE J1752--223 was discovered on 2009 October 23 by RXTE (\citealt{Markwardt2009}). The properties of the X-ray spectrum and the lack of pulsations in the X-ray band suggested that the source was a black-hole candidate (\citealt{Markwardt2009b}), conclusion supported by \cite{Munoz-Darias2010}. 

Since XTE J1752-223 has not shown absorption dips nor eclipses, we can assume an upper limit on the inclination of $\sim$70$^{\circ}$. \cite{Miller-Jones2011} reported the detection of compact radio emission from the core of  XTE J1752-223, which can be associated to the relativistic ejections from the source. Assuming the radio jet as perpendicular to the orbital plane, the inclination of  XTE J1752-223 must be lower than $\sim$49$^{\circ}$. 

\subsection{XTE J1859+226}

XTE J1859+226 was discovered during its 1999 outburst (\citealt{Wood1999}) and its X-ray properties allowed to classify it as a BH candidate. \cite{Corral-Santana2011} performed photometry of the source and found a mass function significantly smaller than the ones previously estimated (see e.g. \citealt{Filippenko2001}), but still consistent with the presence of a black hole.

The lack of eclipses allowed \cite{Corral-Santana2011} to set an upper limit to the source inclination to 70$^{\circ}$. Assuming no contribution from the accretion disk to the optical emission, an inclination of at least 60$^{\circ}$ is needed to reproduce the large modulations observed in the light-curve. A disk contribution to the optical light-curve equal to 25\% would result in an inclination angle of $\sim$70$^{\circ}$. This suggest that XTE J1859+226 might be at intermediate inclination with respect to the line of sight.
This is confirmed by the spectral properties of the source. Despite the lack of clear absorption dips, \cite{Munoz-Darias2013} noted that XTE J1859+27 showed the characteristic flux spikes in its hardness intensity diagrams (HIDs) and lightcurves, typical of high-inclination sources.

\subsection{XTE J1550--564}

XTE J1550--564 was discovered on 1998 (\citealt{Smith1998}) by RXTE. \cite{Orosz2011} performed moderate-resolution optical spectroscopy and near-infrared photometry of the source and were able to determine the mass function for the binary system. \cite{Orosz2011} fitted the light-curves from different wavebands obtained over a 8 years of observations (2001--2009) with a number of models, combining different subsets of data and assuming variable disk-contributions to the optical emission and obtaining different values for the orbital inclination. Although the fit to the light-curve from a small subset of data (collected in 2006-2007) yielded an inclination of $\sim$57$^{\circ}$, the best fit to the overall optical data (considered more reliable by the authors) corresponds to an inclination of (74.69 $\pm$ 3.8)$^{\circ}$.

\subsection{4U 1630--47}

4U 1630-47 has shown outbursts every $\sim$600 days since at least 1969 (\citealt{Kuulkers1997}). Even if the nature of the compact object is still to be confirmed, the X-ray spectral and timing properties of 4U 1630-47 point to a BH.  

\cite{Kuulkers1998} discovered absorption dips of a typical duration of a few minutes in the X-ray light-curve of 4U 1630-47, but no eclipses, that allowed these authors to constrain the inclination of the source between 65$^{\circ}$ and 75$^{\circ}$. The lack of other significant information does not allow to place any other constraints on the source inclination.

\subsection{GRO J1655--40}

GRO J1655-40 was discovered in 1994 by the \textit{Compton Gamma-Ray Observatory} (\citealt{Zhang1994a}). Since this system has a relatively luminous F companion star, the quiescent light-curve is dominated by star light at an unusual extent and, as a result, an unusually precise determination of the orbital parameters is
possible. 

\citealt{Orosz1997} reported a first inclination measurement of 69.50 $\pm$0.08. Later works, based on larger datasets, reported refined inclination measurements, all within the 63$^{\circ}$--75$^{\circ}$ range (\citealt{vanderHooft1998}, \citealt{Shahbaz1999}). \cite{Greene2001} performed multi-wavelenght photometry of GRO J1655-40, obtaining an inclination of (70.2$\pm$1.9)$^{\circ}$. The most recent fit to the  multi-wavelength light-curve of GRO J1655-40 is reported in \cite{Beer2002}, which obtained an inclination of (68.65$\pm$1.5)$^{\circ}$.

\subsection{H1743--322}

First detected during an outburst in 1977 (\citealt{Kaluzienski1977}), H1743-322  was rediscovered as a new source in 2003 by INTEGRAL (\citealt{Revnivtsev2003}) and only subsequently associated with the previously
known source (\citealt{Markwardt2003b}). Since that time, several smaller outbursts of the source have been observed. 
From its similarities with the dynamically confirmed BH XTE J1550-564, H1743-322 was classified as a BH candidate (see \citealt{McClintock2009a}), although no mass function has yet been reported.

\cite{Homan2005b} and \cite{Miller2006} reported observations of absorption dips from this source, appeared only after years of monitoring of the source. This indicates that the orbital inclination is relatively high, possibly on the order of $\sim$80$^{\circ}$. \cite{Steiner2012c} determined the inclination of the radio jet with respect to the line of sight to be (75$\pm$3)$^{\circ}$, which confirms the high inclination of the source if we assume that the disk and the jet axes are (almost) aligned.

\subsection{MAXI J1659--152}

MAXI J1659−152 was discovered in September 2010 by \textit{Swift}/BAT. The source was initially designated as a gamma ray burst and subsequently classified as an X-ray transient thanks to MAXI data (\citealt{Negoro2010}). 

\cite{Kennea2010} reported frequent intensity drops in the X-ray lightcurve, possibly attributed to absorption dips or partial eclipses. \cite{Kuulkers2010} and \citealt{Kuulkers2013} confirmed the detection of absorption disp and established an orbital period of $\sim$ 2.42h for the source. The presence of the absorption dips also allowed these authors to constrain the inclination of the orbital plane of MAXI J1659-152 between 65 and 80$^{\circ}$.

\subsection{MAXI J1543--564}

MAXI J1543-564 was discovered by MAXI in 2011. The spectral and timing properties of the sources allowed to classify the compact object in the system as a BH candidate \citealt{Stiele2011}.

No mass function has been reported for this source and no information are available to allow to measure or constrain the inclination of this source. Hints pointing to intermediate inclination can be found in the typical spikes that can be observed in the HID and light-curve of the system, reported by \cite{Stiele2011} and discussed by \cite{Munoz-Darias2013}. However, since no absorption dips nor eclipses have been observed (yet), it is not possible to further constrain the inclination of the source.

\subsection{XTE J1748--288}

XTE J1748-288 was discovered on 1998 by the RXTE/ASM (\citealt{Smith1998a}). An optically thin radio counterpart was discovered by \cite{Hjellming1998a} and subsequently the source was resolved by VLA (\citealt{Rupen1998}). Follow up observations confirmed the detection of a radio jet, making XTE J1748-288 one of the  few sources displaying apparent superluminal motion (\citealt{Hjellming1998}).

\cite{Revnivtsev2000} reported a sharp dip in the 15–30 keV emission in the RXTE/PCA data, while  \citealt{Kotani2000} derived an inclination of the jet of $<$70$^{\circ}$ from proper motion. These information suggest that XTE J1748-288 is seen at high inclination, even though the absorption nature of the dips observed by \cite{Revnivtsev2000} is to be confirmed.

\bibliographystyle{mn2e.bst}
\bibliography{biblio.bib} 

\begin{thebibliography}{}

\bibitem[\protect\citeauthoryear{{Altamirano}, {Belloni}, {Linares}, {van der
  Klis}, {Wijnands}, {Curran}, {Kalamkar}, {Stiele}, {Motta},
  {Mu{\~n}oz-Darias}, {Casella} \& {Krimm}}{{Altamirano}
  et~al.}{2011}]{Altamirano2011}
{Altamirano} D.,  {Belloni} T.,  {Linares} M.,  {van der Klis} M.,  {Wijnands}
  R.,  {Curran} P.~A.,  {Kalamkar} M.,  {Stiele} H.,  {Motta} S.,
  {Mu{\~n}oz-Darias} T.,  {Casella} P.,    {Krimm} H.,  2011, \apjl, 742, L17

\bibitem[\protect\citeauthoryear{{Antonucci}}{{Antonucci}}{1993}]{Antonucci1993}
{Antonucci} R.,  1993, \araa, 31, 473

\bibitem[\protect\citeauthoryear{{Armitage} \& {Livio}}{{Armitage} \&
  {Livio}}{1998}]{Armitage1998}
{Armitage} P.~J.,  {Livio} M.,  1998, \apj, 493, 898

\bibitem[\protect\citeauthoryear{{Bayless}, {Robinson}, {Hynes}, {Ashcraft} \&
  {Cornell}}{{Bayless} et~al.}{2010}]{Bayless2010}
{Bayless} A.~J.,  {Robinson} E.~L.,  {Hynes} R.~I.,  {Ashcraft} T.~A.,
  {Cornell} M.~E.,  2010, \apj, 709, 251

\bibitem[\protect\citeauthoryear{{Beer} \& {Podsiadlowski}}{{Beer} \&
  {Podsiadlowski}}{2002}]{Beer2002}
{Beer} M.~E.,  {Podsiadlowski} P.,  2002, \mnras, 331, 351

\bibitem[\protect\citeauthoryear{{Belloni} \& {Hasinger}}{{Belloni} \&
  {Hasinger}}{1990}]{Belloni1990}
{Belloni} T.,  {Hasinger} G.,  1990, \aap, 230, 103

\bibitem[\protect\citeauthoryear{{Belloni}, {Homan}, {Motta}, {Ratti} \&
  {M{\'e}ndez}}{{Belloni} et~al.}{2007}]{Belloni2007}
{Belloni} T.,  {Homan} J.,  {Motta} S.,  {Ratti} E.,    {M{\'e}ndez} M.,  2007,
  \mnras, 379, 247

\bibitem[\protect\citeauthoryear{{Belloni}, {Psaltis} \& {van der
  Klis}}{{Belloni} et~al.}{2002}]{Belloni2002}
{Belloni} T.,  {Psaltis} D.,    {van der Klis} M.,  2002, \apj, 572, 392

\bibitem[\protect\citeauthoryear{{Belloni}, {Sanna} \& {M{\'e}ndez}}{{Belloni}
  et~al.}{2012}]{Belloni2012}
{Belloni} T.~M.,  {Sanna} A.,    {M{\'e}ndez} M.,  2012, \mnras, 426, 1701

\bibitem[\protect\citeauthoryear{{Bianchi}, {Maiolino} \& {Risaliti}}{{Bianchi}
  et~al.}{2012}]{Bianchi2012}
{Bianchi} S.,  {Maiolino} R.,    {Risaliti} G.,  2012, Advances in Astronomy,
  2012

\bibitem[\protect\citeauthoryear{{Cabanac}, {Henri}, {Petrucci}, {Malzac},
  {Ferreira} \& {Belloni}}{{Cabanac} et~al.}{2010}]{Cabanac2010}
{Cabanac} C.,  {Henri} G.,  {Petrucci} P.-O.,  {Malzac} J.,  {Ferreira} J.,
  {Belloni} T.~M.,  2010, \mnras, 404, 738

\bibitem[\protect\citeauthoryear{{Casares} \& {Jonker}}{{Casares} \&
  {Jonker}}{2014}]{Casares2014}
{Casares} J.,  {Jonker} P.~G.,  2014, \ssr, 183, 223

\bibitem[\protect\citeauthoryear{{Casella}, {Belloni} \& {Stella}}{{Casella}
  et~al.}{2005}]{Casella2005}
{Casella} P.,  {Belloni} T.,    {Stella} L.,  2005, \apj, 629, 403

\bibitem[\protect\citeauthoryear{{Corbel}, {Kaaret}, {Fender}, {Tzioumis},
  {Tomsick} \& {Orosz}}{{Corbel} et~al.}{2005}]{Corbel2005}
{Corbel} S.,  {Kaaret} P.,  {Fender} R.~P.,  {Tzioumis} A.~K.,  {Tomsick}
  J.~A.,    {Orosz} J.~A.,  2005, \apj, 632, 504

\bibitem[\protect\citeauthoryear{{Corral-Santana}, {Casares},
  {Mu{\~n}oz-Darias}, {Rodr{\'{\i}}guez-Gil}, {Shahbaz}, {Torres}, {Zurita} \&
  {Tyndall}}{{Corral-Santana} et~al.}{2013}]{Corral-Santana2013}
{Corral-Santana} J.~M.,  {Casares} J.,  {Mu{\~n}oz-Darias} T.,
  {Rodr{\'{\i}}guez-Gil} P.,  {Shahbaz} T.,  {Torres} M.~A.~P.,  {Zurita} C.,
   {Tyndall} A.~A.,  2013, Science, 339, 1048

\bibitem[\protect\citeauthoryear{{Corral-Santana}, {Casares}, {Shahbaz},
  {Zurita}, {Mart{\'{\i}}nez-Pais} \& {Rodr{\'{\i}}guez-Gil}}{{Corral-Santana}
  et~al.}{2011}]{Corral-Santana2011}
{Corral-Santana} J.~M.,  {Casares} J.,  {Shahbaz} T.,  {Zurita} C.,
  {Mart{\'{\i}}nez-Pais} I.~G.,    {Rodr{\'{\i}}guez-Gil} P.,  2011, \mnras,
  413, L15

\bibitem[\protect\citeauthoryear{{Cowley}, {Schmidtke}, {Hutchings} \&
  {Crampton}}{{Cowley} et~al.}{2002}]{Cowley2002}
{Cowley} A.~P.,  {Schmidtke} P.~C.,  {Hutchings} J.~B.,    {Crampton} D.,
  2002, \apj, 123, 1741

\bibitem[\protect\citeauthoryear{{D'A{\`i}}, {Iaria}, {Di Salvo}, {Riggio},
  {Burderi} \& {Robba}}{{D'A{\`i}} et~al.}{2014}]{DA`i2014}
{D'A{\`i}} A.,  {Iaria} R.,  {Di Salvo} T.,  {Riggio} A.,  {Burderi} L.,
  {Robba} N.~R.,  2014, \aap, 564, A62

\bibitem[\protect\citeauthoryear{{de Jong}, {van Paradijs} \& {Augusteijn}}{{de
  Jong} et~al.}{1996}]{deJong1996}
{de Jong} J.~A.,  {van Paradijs} J.,    {Augusteijn} T.,  1996, \aap, 314, 484

\bibitem[\protect\citeauthoryear{{D{\'{\i}}az Trigo}, {Parmar}, {Boirin},
  {Motch}, {Talavera} \& {Balman}}{{D{\'{\i}}az Trigo}
  et~al.}{2009}]{DiazTrigo2009}
{D{\'{\i}}az Trigo} M.,  {Parmar} A.~N.,  {Boirin} L.,  {Motch} C.,  {Talavera}
  A.,    {Balman} S.,  2009, \aap, 493, 145

\bibitem[\protect\citeauthoryear{{Esin}, {McClintock} \& {Narayan}}{{Esin}
  et~al.}{1997}]{Esin1997}
{Esin} A.~A.,  {McClintock} J.~E.,    {Narayan} R.,  1997, \apj, 489, 865

\bibitem[\protect\citeauthoryear{{Fender} \& {Belloni}}{{Fender} \&
  {Belloni}}{2004}]{Fender2004a}
{Fender} R.,  {Belloni} T.,  2004, \araa, 42, 317

\bibitem[\protect\citeauthoryear{{Fender}, {Belloni} \& {Gallo}}{{Fender}
  et~al.}{2004}]{Fender2004}
{Fender} R.~P.,  {Belloni} T.~M.,    {Gallo} E.,  2004, \mnras, 355, 1105

\bibitem[\protect\citeauthoryear{{Fender}, {Homan} \& {Belloni}}{{Fender}
  et~al.}{2009}]{Fender2009}
{Fender} R.~P.,  {Homan} J.,    {Belloni} T.~M.,  2009, \mnras, 396, 1370

\bibitem[\protect\citeauthoryear{{Filippenko} \& {Chornock}}{{Filippenko} \&
  {Chornock}}{2001}]{Filippenko2001}
{Filippenko} A.~V.,  {Chornock} R.,  2001, \iaucirc, 7644, 2

\bibitem[\protect\citeauthoryear{{Foulkes}, {Haswell} \& {Murray}}{{Foulkes}
  et~al.}{2010}]{Foulkes2010}
{Foulkes} S.~B.,  {Haswell} C.~A.,    {Murray} J.~R.,  2010, \mnras, 401, 1275

\bibitem[\protect\citeauthoryear{{Frank}, {King} \& {Lasota}}{{Frank}
  et~al.}{1987}]{Frank1987}
{Frank} J.,  {King} A.~R.,    {Lasota} J.-P.,  1987, \aap, 178, 137

\bibitem[\protect\citeauthoryear{{Gallo}, {Fender} \& {Pooley}}{{Gallo}
  et~al.}{2003}]{Gallo2003}
{Gallo} E.,  {Fender} R.~P.,    {Pooley} G.~G.,  2003, \mnras, 344, 60

\bibitem[\protect\citeauthoryear{{Galloway}}{{Galloway}}{2012}]{Galloway2012}
{Galloway} D.~K.,  2012, The Astronomer's Telegram, 4014, 1

\bibitem[\protect\citeauthoryear{{Gierli{\'n}ski}, {Middleton}, {Ward} \&
  {Done}}{{Gierli{\'n}ski} et~al.}{2008}]{Gierlinski2008}
{Gierli{\'n}ski} M.,  {Middleton} M.,  {Ward} M.,    {Done} C.,  2008, \nat,
  455, 369

\bibitem[\protect\citeauthoryear{{Gierli{\'n}ski} \&
  {Zdziarski}}{{Gierli{\'n}ski} \& {Zdziarski}}{2005}]{Gierli'nski2005}
{Gierli{\'n}ski} M.,  {Zdziarski} A.~A.,  2005, \mnras, 363, 1349

\bibitem[\protect\citeauthoryear{{Greene}, {Bailyn} \& {Orosz}}{{Greene}
  et~al.}{2001}]{Greene2001}
{Greene} J.,  {Bailyn} C.~D.,    {Orosz} J.~A.,  2001, \apj, 554, 1290

\bibitem[\protect\citeauthoryear{{Heil}, {Uttley} \& {Klein-Wolt}}{{Heil}
  et~al.}{2014}]{Heil2014}
{Heil} L.~M.,  {Uttley} P.,    {Klein-Wolt} M.,  2014, ArXiv e-prints

\bibitem[\protect\citeauthoryear{{Hjellming}, {Rupen}, {Ghigo}, {Waltman},
  {Mioduszewski}, {Fender}, {Stappers}, {Wieringa}, {Wark} \&
  {Green}}{{Hjellming} et~al.}{1998}]{Hjellming1998a}
{Hjellming} R.~M.,  {Rupen} M.~P.,  {Ghigo} F.,  {Waltman} E.~B.,
  {Mioduszewski} A.~J.,  {Fender} R.~P.,  {Stappers} B.~W.,  {Wieringa} M.,
  {Wark} R.,    {Green} D.~W.~E.,  1998, \iaucirc, 6937, 1

\bibitem[\protect\citeauthoryear{{Hjellming}, {Rupen}, {Mioduszewski}, {Smith},
  {Harmon}, {Waltman}, {Ghigo} \& {Pooley}}{{Hjellming}
  et~al.}{1998}]{Hjellming1998}
{Hjellming} R.~M.,  {Rupen} M.~P.,  {Mioduszewski} A.~J.,  {Smith} D.~A.,
  {Harmon} B.~A.,  {Waltman} E.~B.,  {Ghigo} F.~D.,    {Pooley} G.~G.,  1998,
  in American Astronomical Society, 193rd AAS Meeting, \#103.08; Bulletin of
  the American Astronomical Society, Vol. 30, p.1405</ALTJOURNAL> Vol.~30 of
  Bulletin of the American Astronomical Society, {Radio and X-ray Observations
  of the New Relativistic Jet X-ray Transient XTE J1748-288}.
p. 103.08

\bibitem[\protect\citeauthoryear{{Homan}, {Buxton}, {Markoff}, {Bailyn},
  {Nespoli} \& {Belloni}}{{Homan} et~al.}{2005}]{Homan2005}
{Homan} J.,  {Buxton} M.,  {Markoff} S.,  {Bailyn} C.~D.,  {Nespoli} E.,
  {Belloni} T.,  2005, \apj, 624, 295

\bibitem[\protect\citeauthoryear{{Homan}, {Miller}, {Wijnands}, {van der Klis},
  {Belloni}, {Steeghs} \& {Lewin}}{{Homan} et~al.}{2005b}]{Homan2005b}
{Homan} J.,  {Miller} J.~M.,  {Wijnands} R.,  {van der Klis} M.,  {Belloni} T.,
   {Steeghs} D.,    {Lewin} W.~H.~G.,  2005b, \apj, 623, 383

\bibitem[\protect\citeauthoryear{{Homan}, {van der Klis}, {Jonker}, {Wijnands},
  {Kuulkers}, {M{\'e}ndez} \& {Lewin}}{{Homan} et~al.}{2002}]{Homan2002a}
{Homan} J.,  {van der Klis} M.,  {Jonker} P.~G.,  {Wijnands} R.,  {Kuulkers}
  E.,  {M{\'e}ndez} M.,    {Lewin} W.~H.~G.,  2002, \apj, 568, 878

\bibitem[\protect\citeauthoryear{{Homan}, {Wijnands}, {van der Klis},
  {Belloni}, {van Paradijs}, {Klein-Wolt}, {Fender} \& {M{\'e}ndez}}{{Homan}
  et~al.}{2001}]{Homan2001}
{Homan} J.,  {Wijnands} R.,  {van der Klis} M.,  {Belloni} T.,  {van Paradijs}
  J.,  {Klein-Wolt} M.,  {Fender} R.,    {M{\'e}ndez} M.,  2001, \apjs, 132,
  377

\bibitem[\protect\citeauthoryear{{Ingram} \& {Done}}{{Ingram} \&
  {Done}}{2011}]{Ingram2011}
{Ingram} A.,  {Done} C.,  2011, \mnras, 415, 2323

\bibitem[\protect\citeauthoryear{{Ingram} \& {Done}}{{Ingram} \&
  {Done}}{2012}]{Ingram2012}
{Ingram} A.,  {Done} C.,  2012, \mnras, 419, 2369

\bibitem[\protect\citeauthoryear{{Ingram}, {Done} \& {Fragile}}{{Ingram}
  et~al.}{2009}]{Ingram2009}
{Ingram} A.,  {Done} C.,    {Fragile} P.~C.,  2009, \mnras, 397, L101

\bibitem[\protect\citeauthoryear{{Ingram} \& {van der Klis}}{{Ingram} \& {van
  der Klis}}{2013}]{Ingram2013}
{Ingram} A.,  {van der Klis} M.~v.~d.,  2013, \mnras, 434, 1476

\bibitem[\protect\citeauthoryear{{Kaluzienski} \& {Holt}}{{Kaluzienski} \&
  {Holt}}{1977}]{Kaluzienski1977}
{Kaluzienski} L.~J.,  {Holt} S.~S.,  1977, \iaucirc, 3099, 1

\bibitem[\protect\citeauthoryear{{Kennea}, {Krimm}, {Mangano}, {Curran},
  {Romano}, {Evans} \& {Burrows}}{{Kennea} et~al.}{2010}]{Kennea2010}
{Kennea} J.~A.,  {Krimm} H.,  {Mangano} V.,  {Curran} P.,  {Romano} P.,
  {Evans} P.,    {Burrows} D.~N.,  2010, The Astronomer's Telegram, 2877, 1

\bibitem[\protect\citeauthoryear{{Kotani}, {Kawai}, {Nagase}, {Namiki},
  {Sakano}, {Takeshima}, {Ueda}, {Yamaoka} \& {Hjellming}}{{Kotani}
  et~al.}{2000}]{Kotani2000}
{Kotani} T.,  {Kawai} N.,  {Nagase} F.,  {Namiki} M.,  {Sakano} M.,
  {Takeshima} T.,  {Ueda} Y.,  {Yamaoka} K.,    {Hjellming} R.~M.,  2000,
  \apjl, 543, L133

\bibitem[\protect\citeauthoryear{{Kuulkers}, {Ibarra}, {Pollock}, {Parmar},
  {Chenevez}, {Kouveliotou}, {van der Horst}, {Paragi}, {Granot} \&
  {Taylor}}{{Kuulkers} et~al.}{2010}]{Kuulkers2010}
{Kuulkers} E.,  {Ibarra} A.,  {Pollock} A.,  {Parmar} A.,  {Chenevez} J.,
  {Kouveliotou} C.,  {van der Horst} A.~J.,  {Paragi} Z.,  {Granot} J.,
  {Taylor} G.~B.,  2010, The Astronomer's Telegram, 2912, 1

\bibitem[\protect\citeauthoryear{{Kuulkers}, {in't Zand}, {Cornelisse},
  {Heise}, {Kong}, {Charles}, {Bazzano}, {Cocchi}, {Natalucci} \&
  {Ubertini}}{{Kuulkers} et~al.}{2000}]{Kuulkers2000}
{Kuulkers} E.,  {in't Zand} J.~J.~M.,  {Cornelisse} R.,  {Heise} J.,  {Kong}
  A.~K.~H.,  {Charles} P.~A.,  {Bazzano} A.,  {Cocchi} M.,  {Natalucci} L.,
  {Ubertini} P.,  2000, \aap, 358, 993

\bibitem[\protect\citeauthoryear{{Kuulkers}, {Kouveliotou}, {Belloni}, {Cadolle
  Bel}, {Chenevez}, {D{\'{\i}}az Trigo}, {Homan}, {Ibarra}, {Kennea},
  {Mu{\~n}oz-Darias}, {Ness}, {Parmar}, {Pollock}, {van den Heuvel} \& {van der
  Horst}}{{Kuulkers} et~al.}{2013}]{Kuulkers2013}
{Kuulkers} E.,  {Kouveliotou} C.,  {Belloni} T.,  {Cadolle Bel} M.,  {Chenevez}
  J.,  {D{\'{\i}}az Trigo} M.,  {Homan} J.,  {Ibarra} A.,  {Kennea} J.~A.,
  {Mu{\~n}oz-Darias} T.,  {Ness} J.-U.,  {Parmar} A.~N.,  {Pollock} A.~M.~T.,
  {van den Heuvel} E.~P.~J.,    {van der Horst} A.~J.,  2013, \aap, 552, A32

\bibitem[\protect\citeauthoryear{{Kuulkers}, {Parmar}, {Kitamoto}, {Cominsky}
  \& {Sood}}{{Kuulkers} et~al.}{1997}]{Kuulkers1997}
{Kuulkers} E.,  {Parmar} A.~N.,  {Kitamoto} S.,  {Cominsky} L.~R.,    {Sood}
  R.~K.,  1997, \mnras, 291, 81

\bibitem[\protect\citeauthoryear{{Kuulkers}, {Wijnands}, {Belloni}, {Mendez},
  {van der Klis} \& {van Paradijs}}{{Kuulkers} et~al.}{1998}]{Kuulkers1998}
{Kuulkers} E.,  {Wijnands} R.,  {Belloni} T.,  {Mendez} M.,  {van der Klis} M.,
     {van Paradijs} J.,  1998, \apj, 494, 753

\bibitem[\protect\citeauthoryear{{Lamb} \& {Miller}}{{Lamb} \&
  {Miller}}{2001}]{Lamb2001}
{Lamb} F.~K.,  {Miller} M.~C.,  2001, \apj, 554, 1210

\bibitem[\protect\citeauthoryear{{Leahy}, {Elsner} \& {Weisskopf}}{{Leahy}
  et~al.}{1983}]{Leahy1983}
{Leahy} D.~A.,  {Elsner} R.~F.,    {Weisskopf} M.~C.,  1983, \apj, 272, 256

\bibitem[\protect\citeauthoryear{{Malzac}}{{Malzac}}{2013}]{Malzac2013}
{Malzac} J.,  2013, \mnras, 429, L20

\bibitem[\protect\citeauthoryear{Mann \& Whitney}{Mann \&
  Whitney}{1947}]{Mann1947}
Mann H.~B.,  Whitney D.~R.,  1947, The Annals of Mathematical Statistics, 18,
  50

\bibitem[\protect\citeauthoryear{{Markert}, {Canizares}, {Clark}, {Lewin},
  {Schnopper} \& {Sprott}}{{Markert} et~al.}{1973}]{Markert1973}
{Markert} T.~H.,  {Canizares} C.~R.,  {Clark} G.~W.,  {Lewin} W.~H.~G.,
  {Schnopper} H.~W.,    {Sprott} G.~F.,  1973, \apjl, 184, L67+

\bibitem[\protect\citeauthoryear{{Markwardt}, {Barthelmy}, {Evans} \&
  {Swank}}{{Markwardt} et~al.}{2009}]{Markwardt2009b}
{Markwardt} C.~B.,  {Barthelmy} S.~D.,  {Evans} P.~A.,    {Swank} J.~H.,  2009,
  The Astronomer's Telegram, 2261, 1

\bibitem[\protect\citeauthoryear{{Markwardt} \& {Swank}}{{Markwardt} \&
  {Swank}}{2003}]{Markwardt2003b}
{Markwardt} C.~B.,  {Swank} J.~H.,  2003, The Astronomer's Telegram, 133, 1

\bibitem[\protect\citeauthoryear{{Markwardt}, {Swank}, {Barthelmy},
  {Baumgartner}, {Burrows}, {Evans}, {Holland}, {Hoversten} \&
  {Page}}{{Markwardt} et~al.}{2009}]{Markwardt2009}
{Markwardt} C.~B.,  {Swank} J.~H.,  {Barthelmy} S.~D.,  {Baumgartner} W.~H.,
  {Burrows} D.~N.,  {Evans} P.~A.,  {Holland} S.~T.,  {Hoversten} E.~A.,
  {Page} K.~L.,  2009, The Astronomer's Telegram, 2258, 1

\bibitem[\protect\citeauthoryear{{Matilsky}, {Giacconi}, {Gursky}, {Kellogg} \&
  {Tananbaum}}{{Matilsky} et~al.}{1972}]{Matilsky1972}
{Matilsky} T.~A.,  {Giacconi} R.,  {Gursky} H.,  {Kellogg} E.~M.,
  {Tananbaum} H.~D.,  1972, \apjl, 174, L53

\bibitem[\protect\citeauthoryear{{McClintock}, {Remillard}, {Rupen}, {Torres},
  {Steeghs}, {Levine} \& {Orosz}}{{McClintock} et~al.}{2009}]{McClintock2009a}
{McClintock} J.~E.,  {Remillard} R.~A.,  {Rupen} M.~P.,  {Torres} M.~A.~P.,
  {Steeghs} D.,  {Levine} A.~M.,    {Orosz} J.~A.,  2009, \apj, 698, 1398

\bibitem[\protect\citeauthoryear{{Middleton} \& {Done}}{{Middleton} \&
  {Done}}{2010}]{Middleton2010}
{Middleton} M.,  {Done} C.,  2010, \mnras, 403, 9

\bibitem[\protect\citeauthoryear{{Miller}, {Raymond}, {Homan}, {Fabian},
  {Steeghs}, {Wijnands}, {Rupen}, {Charles}, {van der Klis} \&
  {Lewin}}{{Miller} et~al.}{2006}]{Miller2006}
{Miller} J.~M.,  {Raymond} J.,  {Homan} J.,  {Fabian} A.~C.,  {Steeghs} D.,
  {Wijnands} R.,  {Rupen} M.,  {Charles} P.,  {van der Klis} M.,    {Lewin}
  W.~H.~G.,  2006, \apj, 646, 394

\bibitem[\protect\citeauthoryear{{Miller-Jones}, {Jonker}, {Ratti}, {Torres},
  {Brocksopp}, {Yang} \& {Morrell}}{{Miller-Jones}
  et~al.}{2011}]{Miller-Jones2011}
{Miller-Jones} J.~C.~A.,  {Jonker} P.~G.,  {Ratti} E.~M.,  {Torres} M.~A.~P.,
  {Brocksopp} C.,  {Yang} J.,    {Morrell} N.~I.,  2011, \mnras, 415, 306

\bibitem[\protect\citeauthoryear{{Miller-Jones}, {Sivakoff}, {Altamirano},
  {Coriat}, {Corbel}, {Dhawan}, {Krimm}, {Remillard}, {Rupen}, {Russell},
  {Fender}, {Heinz}, {Kording}, {Maitra} \& et al.}{{Miller-Jones}
  et~al.}{2012}]{Miller-Jones2012}
{Miller-Jones} J.~C.~A.,  {Sivakoff} G.~R.,  {Altamirano} D.,  {Coriat} M.,
  {Corbel} S.,  {Dhawan} V.,  {Krimm} H.~A.,  {Remillard} R.~A.,  {Rupen}
  M.~P.,  {Russell} D.~M.,  {Fender} R.~P.,  {Heinz} S.,  {Kording} E.~G.,
  {Maitra} D.,    et al. M.,  2012, \mnras, 421, 468

\bibitem[\protect\citeauthoryear{{Miyamoto}, {Kimura}, {Kitamoto}, {Dotani} \&
  {Ebisawa}}{{Miyamoto} et~al.}{1991}]{Miyamoto1991a}
{Miyamoto} S.,  {Kimura} K.,  {Kitamoto} S.,  {Dotani} T.,    {Ebisawa} K.,
  1991, \apj, 383, 784

\bibitem[\protect\citeauthoryear{{Miyamoto} \& {Kitamoto}}{{Miyamoto} \&
  {Kitamoto}}{1991}]{Miyamoto1991}
{Miyamoto} S.,  {Kitamoto} S.,  1991, \apj, 374, 741

\bibitem[\protect\citeauthoryear{{Motch}, {Ricketts}, {Page}, {Ilovaisky} \&
  {Chevalier}}{{Motch} et~al.}{1983}]{Motch1983}
{Motch} C.,  {Ricketts} M.~J.,  {Page} C.~G.,  {Ilovaisky} S.~A.,
  {Chevalier} C.,  1983, \aap, 119, 171

\bibitem[\protect\citeauthoryear{{Motta}, {Homan}, {Mu{\~n}oz Darias},
  {Casella}, {Belloni}, {Hiemstra} \& {M{\'e}ndez}}{{Motta}
  et~al.}{2012}]{Motta2012}
{Motta} S.,  {Homan} J.,  {Mu{\~n}oz Darias} T.,  {Casella} P.,  {Belloni}
  T.~M.,  {Hiemstra} B.,    {M{\'e}ndez} M.,  2012, \mnras, 427, 595

\bibitem[\protect\citeauthoryear{{Motta}, {Mu{\~n}oz-Darias}, {Casella},
  {Belloni} \& {Homan}}{{Motta} et~al.}{2011a}]{Motta2011a}
{Motta} S.,  {Mu{\~n}oz-Darias} T.,  {Casella} P.,  {Belloni} T.,    {Homan}
  J.,  2011a, \mnras, 418, 2292

\bibitem[\protect\citeauthoryear{{Motta}, {Mu{\~n}oz-Darias}, {Casella},
  {Belloni} \& {Homan}}{{Motta} et~al.}{2011b}]{Motta2011}
{Motta} S.,  {Mu{\~n}oz-Darias} T.,  {Casella} P.,  {Belloni} T.,    {Homan}
  J.,  2011b, \mnras, 418, 2292

\bibitem[\protect\citeauthoryear{{Motta}, {Belloni}, {Stella},
  {Mu{\~n}oz-Darias} \& {Fender}}{{Motta} et~al.}{2014}]{Motta2014}
{Motta} S.~E.,  {Belloni} T.~M.,  {Stella} L.,  {Mu{\~n}oz-Darias} T.,
  {Fender} R.,  2014, \mnras, 437, 2554

\bibitem[\protect\citeauthoryear{{Motta}, {Mu{\~n}oz-Darias}, {Sanna},
  {Fender}, {Belloni} \& {Stella}}{{Motta} et~al.}{2014a}]{Motta2014a}
{Motta} S.~E.,  {Mu{\~n}oz-Darias} T.,  {Sanna} A.,  {Fender} R.,  {Belloni}
  T.,    {Stella} L.,  2014a, \mnras

\bibitem[\protect\citeauthoryear{{Mu{\~n}oz-Darias}, {Casares} \&
  {Mart{\'{\i}}nez-Pais}}{{Mu{\~n}oz-Darias} et~al.}{2008}]{Munoz-Darias2008}
{Mu{\~n}oz-Darias} T.,  {Casares} J.,    {Mart{\'{\i}}nez-Pais} I.~G.,  2008,
  \mnras, 385, 2205

\bibitem[\protect\citeauthoryear{{Mu{\~n}oz-Darias}, {Coriat}, {Plant},
  {Ponti}, {Fender} \& {Dunn}}{{Mu{\~n}oz-Darias}
  et~al.}{2013}]{Munoz-Darias2013}
{Mu{\~n}oz-Darias} T.,  {Coriat} M.,  {Plant} D.~S.,  {Ponti} G.,  {Fender}
  R.~P.,    {Dunn} R.~J.~H.,  2013, \mnras, 432, 1330

\bibitem[\protect\citeauthoryear{{Mu{\~n}oz-Darias}, {Motta}, {Pawar},
  {Belloni}, {Campana} \& {Bhattacharya}}{{Mu{\~n}oz-Darias}
  et~al.}{2010}]{Munoz-Darias2010}
{Mu{\~n}oz-Darias} T.,  {Motta} S.,  {Pawar} D.,  {Belloni} T.~M.,  {Campana}
  S.,    {Bhattacharya} D.,  2010, \mnras, 404, L94

\bibitem[\protect\citeauthoryear{{Negoro}, {Yamaoka}, {Nakahira}, {Kawasaki},
  {Ueno}, {Tomida}, {Kohama}, {Ishikawa}, {Mihara}, {Nakagawa}, {Sugizaki},
  {Serino}, {Yamamoto}, {Sootome}, {Matsuoka}, {Kawai}, {Morii}, {Sugimori} \&
  {Usui}}{{Negoro} et~al.}{2010}]{Negoro2010}
{Negoro} H.,  {Yamaoka} K.,  {Nakahira} S.,  {Kawasaki} K.,  {Ueno} S.,
  {Tomida} H.,  {Kohama} M.,  {Ishikawa} M.,  {Mihara} T.,  {Nakagawa} Y.~E.,
  {Sugizaki} M.,  {Serino} M.,  {Yamamoto} T.,  {Sootome} T.,  {Matsuoka} M.,
  {Kawai} N.,  {Morii} M.,  {Sugimori} K.,    {Usui} R.,  2010, The
  Astronomer's Telegram, 2873, 1

\bibitem[\protect\citeauthoryear{{Nespoli}, {Belloni}, {Homan}, {Miller},
  {Lewin}, {M{\'e}ndez} \& {van der Klis}}{{Nespoli}
  et~al.}{2003}]{Nespoli2003}
{Nespoli} E.,  {Belloni} T.,  {Homan} J.,  {Miller} J.~M.,  {Lewin} W.~H.~G.,
  {M{\'e}ndez} M.,    {van der Klis} M.,  2003, \aap, 412, 235

\bibitem[\protect\citeauthoryear{{Neustroev}, {Veledina}, {Poutanen},
  {Zharikov}, {Tsygankov}, {Sjoberg} \& {Kajava}}{{Neustroev}
  et~al.}{2014}]{Neustroev2014}
{Neustroev} V.~V.,  {Veledina} A.,  {Poutanen} J.,  {Zharikov} S.~V.,
  {Tsygankov} S.~S.,  {Sjoberg} G.,    {Kajava} J.~J.~E.,  2014, \mnras, 445,
  2424

\bibitem[\protect\citeauthoryear{{Orosz}, {Groot}, {van der Klis},
  {McClintock}, {Garcia}, {Zhao}, {Jain}, {Bailyn} \& {Remillard}}{{Orosz}
  et~al.}{2002}]{Orosz2002}
{Orosz} J.~A.,  {Groot} P.~J.,  {van der Klis} M.,  {McClintock} J.~E.,
  {Garcia} M.~R.,  {Zhao} P.,  {Jain} R.~K.,  {Bailyn} C.~D.,    {Remillard}
  R.~A.,  2002, \apj, 568, 845

\bibitem[\protect\citeauthoryear{{Orosz}, {Jain}, {Bailyn}, {McClintock} \&
  {Remillard}}{{Orosz} et~al.}{1998}]{Orosz1998}
{Orosz} J.~A.,  {Jain} R.~K.,  {Bailyn} C.~D.,  {McClintock} J.~E.,
  {Remillard} R.~A.,  1998, \apj, 499, 375

\bibitem[\protect\citeauthoryear{{Orosz}, {McClintock}, {Remillard} \&
  {Corbel}}{{Orosz} et~al.}{2004}]{Orosz2004}
{Orosz} J.~A.,  {McClintock} J.~E.,  {Remillard} R.~A.,    {Corbel} S.,  2004,
  \apj, 616, 376

\bibitem[\protect\citeauthoryear{{Orosz}, {Remillard}, {Bailyn} \&
  {McClintock}}{{Orosz} et~al.}{1997}]{Orosz1997}
{Orosz} J.~A.,  {Remillard} R.~A.,  {Bailyn} C.~D.,    {McClintock} J.~E.,
  1997, \apjl, 478, L83

\bibitem[\protect\citeauthoryear{{Orosz}, {Steiner}, {McClintock}, {Torres},
  {Remillard}, {Bailyn} \& {Miller}}{{Orosz} et~al.}{2011}]{Orosz2011}
{Orosz} J.~A.,  {Steiner} J.~F.,  {McClintock} J.~E.,  {Torres} M.~A.~P.,
  {Remillard} R.~A.,  {Bailyn} C.~D.,    {Miller} J.~M.,  2011, \apj, 730, 75

\bibitem[\protect\citeauthoryear{{Palmer}, {Barthelmey}, {Cummings}, {Gehrels},
  {Krimm}, {Markwardt}, {Sakamoto} \& {Tueller}}{{Palmer}
  et~al.}{2005}]{Palmer2005}
{Palmer} D.~M.,  {Barthelmey} S.~D.,  {Cummings} J.~R.,  {Gehrels} N.,  {Krimm}
  H.~A.,  {Markwardt} C.~B.,  {Sakamoto} T.,    {Tueller} J.,  2005, The
  Astronomer's Telegram, 546, 1

\bibitem[\protect\citeauthoryear{{Parmar}, {Oosterbroek}, {Boirin} \&
  {Lumb}}{{Parmar} et~al.}{2002}]{Parmar2002}
{Parmar} A.~N.,  {Oosterbroek} T.,  {Boirin} L.,    {Lumb} D.,  2002, \aap,
  386, 910

\bibitem[\protect\citeauthoryear{{Patterson}, {Robinson} \&
  {Nather}}{{Patterson} et~al.}{1977}]{Patterson1977}
{Patterson} J.,  {Robinson} E.~L.,    {Nather} R.~E.,  1977, \apj, 214, 144

\bibitem[\protect\citeauthoryear{{Ponti}, {Fender}, {Begelman}, {Dunn},
  {Neilsen} \& {Coriat}}{{Ponti} et~al.}{2012}]{Ponti2012}
{Ponti} G.,  {Fender} R.~P.,  {Begelman} M.~C.,  {Dunn} R.~J.~H.,  {Neilsen}
  J.,    {Coriat} M.,  2012, \mnras, 422, L11

\bibitem[\protect\citeauthoryear{{Poutanen}, {Krolik} \& {Ryde}}{{Poutanen}
  et~al.}{1997}]{Poutanen1997}
{Poutanen} J.,  {Krolik} J.~H.,    {Ryde} F.,  1997, \mnras, 292, L21

\bibitem[\protect\citeauthoryear{{R Development Core Team}}{{R Development Core
  Team}}{2011}]{R_manual}
{R Development Core Team} 2011, R: A Language and Environment for Statistical
  Computing.
R Foundation for Statistical Computing, Vienna, Austria

\bibitem[\protect\citeauthoryear{{Remillard}}{{Remillard}}{2001}]{Remillard2001}
{Remillard} R.,  2001, \iaucirc, 7707, 1

\bibitem[\protect\citeauthoryear{{Remillard}, {Muno}, {McClintock} \&
  {Orosz}}{{Remillard} et~al.}{2002}]{Remillard2002}
{Remillard} R.~A.,  {Muno} M.~P.,  {McClintock} J.~E.,    {Orosz} J.~A.,  2002,
  \apj, 580, 1030

\bibitem[\protect\citeauthoryear{{Revnivtsev}, {Chernyakova}, {Capitanio},
  {Westergaard}, {Shoenfelder}, {Gehrels} \& {Winkler}}{{Revnivtsev}
  et~al.}{2003}]{Revnivtsev2003}
{Revnivtsev} M.,  {Chernyakova} M.,  {Capitanio} F.,  {Westergaard} N.~J.,
  {Shoenfelder} V.,  {Gehrels} N.,    {Winkler} C.,  2003, The Astronomer's
  Telegram, 132, 1

\bibitem[\protect\citeauthoryear{{Revnivtsev}, {Trudolyubov} \&
  {Borozdin}}{{Revnivtsev} et~al.}{2000}]{Revnivtsev2000}
{Revnivtsev} M.~G.,  {Trudolyubov} S.~P.,    {Borozdin} K.~N.,  2000, \mnras,
  312, 151

\bibitem[\protect\citeauthoryear{{Risaliti}, {Elvis} \& {Nicastro}}{{Risaliti}
  et~al.}{2002}]{Risaliti2002}
{Risaliti} G.,  {Elvis} M.,    {Nicastro} F.,  2002, \apj, 571, 234

\bibitem[\protect\citeauthoryear{{Robinson}, {Welsh} \& {Young}}{{Robinson}
  et~al.}{2001}]{Robinson2001}
{Robinson} E.~L.,  {Welsh} W.~F.,    {Young} P.,  2001, X-ray Astronomy:
  Stellar Endpoints, AGN, and the Diffuse X-ray Background, 599, 902

\bibitem[\protect\citeauthoryear{{Rupen}, {Hjellming} \&
  {Mioduszewski}}{{Rupen} et~al.}{1998}]{Rupen1998}
{Rupen} M.~P.,  {Hjellming} R.~M.,    {Mioduszewski} A.~J.,  1998, \iaucirc,
  6938, 2

\bibitem[\protect\citeauthoryear{{Schnittman}, {Homan} \&
  {Miller}}{{Schnittman} et~al.}{2006}]{Schnittman2006}
{Schnittman} J.~D.,  {Homan} J.,    {Miller} J.~M.,  2006, \apj, 642, 420

\bibitem[\protect\citeauthoryear{{Shahbaz}, {van der Hooft}, {Casares},
  {Charles} \& {van Paradijs}}{{Shahbaz} et~al.}{1999}]{Shahbaz1999}
{Shahbaz} T.,  {van der Hooft} F.,  {Casares} J.,  {Charles} P.~A.,    {van
  Paradijs} J.,  1999, \mnras, 306, 89

\bibitem[\protect\citeauthoryear{{Shakura} \& {Sunyaev}}{{Shakura} \&
  {Sunyaev}}{1973}]{Shakura1973}
{Shakura} N.~I.,  {Sunyaev} R.~A.,  1973, \aap, 24, 337

\bibitem[\protect\citeauthoryear{{Smale} \& {Wachter}}{{Smale} \&
  {Wachter}}{1999}]{Smale1999}
{Smale} A.~P.,  {Wachter} S.,  1999, \apj, 527, 341

\bibitem[\protect\citeauthoryear{{Smith}}{{Smith}}{1998}]{Smith1998}
{Smith} D.~A.,  1998, \iaucirc, 7008, 1

\bibitem[\protect\citeauthoryear{{Smith}, {Levine} \& {Wood}}{{Smith}
  et~al.}{1998}]{Smith1998a}
{Smith} D.~A.,  {Levine} A.,    {Wood} A.,  1998, \iaucirc, 6932, 1

\bibitem[\protect\citeauthoryear{{Soleri}, {Belloni} \& {Casella}}{{Soleri}
  et~al.}{2008}]{Soleri2008}
{Soleri} P.,  {Belloni} T.,    {Casella} P.,  2008, \mnras, 383, 1089

\bibitem[\protect\citeauthoryear{{Soleri} \& {Fender}}{{Soleri} \&
  {Fender}}{2011}]{Soleri2011}
{Soleri} P.,  {Fender} R.,  2011, \mnras, 413, 2269

\bibitem[\protect\citeauthoryear{{Soleri}, {Mu{\~n}oz-Darias}, {Motta},
  {Belloni}, {Casella}, {M{\'e}ndez}, {Altamirano}, {Linares}, {Wijnands},
  {Fender} \& {van der Klis}}{{Soleri} et~al.}{2013}]{Soleri2013}
{Soleri} P.,  {Mu{\~n}oz-Darias} T.,  {Motta} S.,  {Belloni} T.,  {Casella} P.,
   {M{\'e}ndez} M.,  {Altamirano} D.,  {Linares} M.,  {Wijnands} R.,  {Fender}
  R.,    {van der Klis} M.,  2013, \mnras, 429, 1244

\bibitem[\protect\citeauthoryear{{Steiner}, {McClintock} \& {Reid}}{{Steiner}
  et~al.}{2012}]{Steiner2012c}
{Steiner} J.~F.,  {McClintock} J.~E.,    {Reid} M.~J.,  2012, \apjl, 745, L7

\bibitem[\protect\citeauthoryear{{Steiner}, {Reis}, {Fabian}, {Remillard},
  {McClintock}, {Gou}, {Cooke}, {Brenneman} \& {Sanders}}{{Steiner}
  et~al.}{2012}]{Steiner2012}
{Steiner} J.~F.,  {Reis} R.~C.,  {Fabian} A.~C.,  {Remillard} R.~A.,
  {McClintock} J.~E.,  {Gou} L.,  {Cooke} R.,  {Brenneman} L.~W.,    {Sanders}
  J.~S.,  2012, \mnras, 427, 2552

\bibitem[\protect\citeauthoryear{{Stella} \& {Vietri}}{{Stella} \&
  {Vietri}}{1999}]{Stella1999a}
{Stella} L.,  {Vietri} M.,  1999, Physical Review Letters, 82, 17

\bibitem[\protect\citeauthoryear{{Stiele}, {Motta}, {Mu{\~n}oz-Darias} \&
  {Belloni}}{{Stiele} et~al.}{2011}]{Stiele2011}
{Stiele} H.,  {Motta} S.,  {Mu{\~n}oz-Darias} T.,    {Belloni} T.~M.,  2011,
  ArXiv e-prints

\bibitem[\protect\citeauthoryear{{Strohmayer} \& {Mushotzky}}{{Strohmayer} \&
  {Mushotzky}}{2003}]{Strohmayer2003a}
{Strohmayer} T.~E.,  {Mushotzky} R.~F.,  2003, \apjl, 586, L61

\bibitem[\protect\citeauthoryear{{Tagger} \& {Pellat}}{{Tagger} \&
  {Pellat}}{1999}]{Tagger1999}
{Tagger} M.,  {Pellat} R.,  1999, \aap, 349, 1003

\bibitem[\protect\citeauthoryear{{Takizawa}, {Dotani}, {Mitsuda}, {Matsuba},
  {Ogawa}, {Aoki}, {Asai}, {Ebisawa}, {Makishima}, {Miyamoto}, {Iga},
  {Vaughan}, {Rutledge} \& {Lewin}}{{Takizawa} et~al.}{1997}]{Takizawa1997}
{Takizawa} M.,  {Dotani} T.,  {Mitsuda} K.,  {Matsuba} E.,  {Ogawa} M.,  {Aoki}
  T.,  {Asai} K.,  {Ebisawa} K.,  {Makishima} K.,  {Miyamoto} S.,  {Iga} S.,
  {Vaughan} B.,  {Rutledge} R.~E.,    {Lewin} W.~H.~G.,  1997, \apj, 489, 272

\bibitem[\protect\citeauthoryear{{Titarchuk} \& {Fiorito}}{{Titarchuk} \&
  {Fiorito}}{2004}]{Titarchuk2004}
{Titarchuk} L.,  {Fiorito} R.,  2004, \apj, 612, 988

\bibitem[\protect\citeauthoryear{{Titarchuk} \& {Osherovich}}{{Titarchuk} \&
  {Osherovich}}{1999}]{Titarchuk1999}
{Titarchuk} L.,  {Osherovich} V.,  1999, \apjl, 518, L95

\bibitem[\protect\citeauthoryear{{Tomsick}, {Lapshov} \& {Kaaret}}{{Tomsick}
  et~al.}{1998}]{Tomsick1998}
{Tomsick} J.~A.,  {Lapshov} I.,    {Kaaret} P.,  1998, \apj, 494, 747

\bibitem[\protect\citeauthoryear{{Urry} \& {Padovani}}{{Urry} \&
  {Padovani}}{1995}]{Urry1995}
{Urry} C.~M.,  {Padovani} P.,  1995, \pasp, 107, 803

\bibitem[\protect\citeauthoryear{{van der Hooft}, {Heemskerk}, {Alberts} \&
  {van Paradijs}}{{van der Hooft} et~al.}{1998}]{vanderHooft1998}
{van der Hooft} F.,  {Heemskerk} M.~H.~M.,  {Alberts} F.,    {van Paradijs} J.,
   1998, \aap, 329, 538

\bibitem[\protect\citeauthoryear{{van der Klis}}{{van der
  Klis}}{1989}]{VdK1989}
{van der Klis} M.,  1989, \araa, 27, 517

\bibitem[\protect\citeauthoryear{{Varni\`ere} \& {Blackman}}{{Varni\`ere} \&
  {Blackman}}{2005}]{Varni`ere2005}
{Varni\`ere} P.,  {Blackman} E.~G.,  2005, \na, 11, 43

\bibitem[\protect\citeauthoryear{{Varni{\`e}re} \& {Tagger}}{{Varni{\`e}re} \&
  {Tagger}}{2002}]{Varni`ere2002}
{Varni{\`e}re} P.,  {Tagger} M.,  2002, \aap, 394, 329

\bibitem[\protect\citeauthoryear{{Varni{\`e}re}, {Tagger} \&
  {Rodriguez}}{{Varni{\`e}re} et~al.}{2012}]{Varni`ere2012}
{Varni{\`e}re} P.,  {Tagger} M.,    {Rodriguez} J.,  2012, \aap, 545, A40

\bibitem[\protect\citeauthoryear{{Veledina}, {Poutanen} \& {Ingram}}{{Veledina}
  et~al.}{2013}]{Veledina2013}
{Veledina} A.,  {Poutanen} J.,    {Ingram} A.,  2013, \apj, 778, 165

\bibitem[\protect\citeauthoryear{{Walter}, {Mason}, {Clarke}, {Halpern},
  {Grindlay}, {Bowyer} \& {Henry}}{{Walter} et~al.}{1982}]{Walter1982}
{Walter} F.~M.,  {Mason} K.~O.,  {Clarke} J.~T.,  {Halpern} J.,  {Grindlay}
  J.~E.,  {Bowyer} S.,    {Henry} J.~P.,  1982, \apjl, 253, L67

\bibitem[\protect\citeauthoryear{{White} \& {Holt}}{{White} \&
  {Holt}}{1982}]{White1982a}
{White} N.~E.,  {Holt} S.~S.,  1982, \apj, 257, 318

\bibitem[\protect\citeauthoryear{{White} \& {Mason}}{{White} \&
  {Mason}}{1985}]{White1985}
{White} N.~E.,  {Mason} K.~O.,  1985, \ssr, 40, 167

\bibitem[\protect\citeauthoryear{{White} \& {Swank}}{{White} \&
  {Swank}}{1982}]{White1982}
{White} N.~E.,  {Swank} J.~H.,  1982, \apjl, 253, L61

\bibitem[\protect\citeauthoryear{{Wijnands}, {Homan} \& {van der
  Klis}}{{Wijnands} et~al.}{1999}]{Wijnands1999}
{Wijnands} R.,  {Homan} J.,    {van der Klis} M.,  1999, \apjl, 526, L33

\bibitem[\protect\citeauthoryear{{Wijnands}, {Miller} \& {Lewin}}{{Wijnands}
  et~al.}{2001}]{Wijnands2001}
{Wijnands} R.,  {Miller} J.~M.,    {Lewin} W.~H.~G.,  2001, \iaucirc, 7715, 2

\bibitem[\protect\citeauthoryear{{Wood}, {Smith}, {Marshall} \& {Swank}}{{Wood}
  et~al.}{1999}]{Wood1999}
{Wood} A.,  {Smith} D.~A.,  {Marshall} F.~E.,    {Swank} J.,  1999, \iaucirc,
  7274, 1

\bibitem[\protect\citeauthoryear{{Zhang}, {Wilson}, {Harmon}, {Fishman},
  {Wilson}, {Paciesas}, {Scott} \& {Rubin}}{{Zhang} et~al.}{1994}]{Zhang1994a}
{Zhang} S.~N.,  {Wilson} C.~A.,  {Harmon} B.~A.,  {Fishman} G.~J.,  {Wilson}
  R.~B.,  {Paciesas} W.~S.,  {Scott} M.,    {Rubin} B.~C.,  1994, \iaucirc,
  6046, 1

\bibitem[\protect\citeauthoryear{{Zhang}, {Jahoda}, {Swank}, {Morgan} \&
  {Giles}}{{Zhang} et~al.}{1995}]{Zhang1995}
{Zhang} W.,  {Jahoda} K.,  {Swank} J.~H.,  {Morgan} E.~H.,    {Giles} A.~B.,
  1995, \apj, 449, 930

\end{thebibliography}
\label{lastpage}
\end{document}